\documentclass[aps,prd,showpacs,floatfix,nofootinbib,amssymb,preprint]{revtex4-1}
\usepackage{graphicx}
\usepackage{slashed}
\usepackage{booktabs}
\usepackage{latexsym}
\usepackage{amsmath}
\usepackage{amsfonts}
\usepackage{amsbsy}
\usepackage{multirow}
\usepackage{dsfont} 

\newcommand{\nn}{\nonumber}
\newcommand{\be}{\begin{equation}}
\newcommand{\ee}{\end{equation}}
\def\al{\alpha}
\def\als{\alpha_{\mathrm{s}}}

\newcommand{\OS}{\mathrm{OS}}
\newcommand{\MS}{\overline{\mathrm{MS}}}
\newcommand{\lQ}{\Lambda_{\mathrm{QCD}}}
\newcommand{\RS}{\mathrm{RS}}

\newcommand{\Eqre}{Eq.~\eqref}

\newcommand{\nmax}{{n_\mathrm{max}}}

\newcommand{\hmu}{\hat{\mu}}
\newcommand{\hnu}{\hat{\nu}}
\newcommand{\Umu}{U_{\mu}}

\begin{document}
\title{
Perturbative expansion of the energy of static sources at large orders
in four-dimensional SU(3) gauge theory
}
\author{Gunnar S.\ Bali}
\email{gunnar.bali@ur.de}
\affiliation{Institut f\"ur Theoretische Physik, Universit\"at
Regensburg, D-93040 Regensburg, Germany}
\affiliation{Tata Institute of Fundamental Research, Homi Bhabha Road, Mumbai 400005, India}
\author{Clemens Bauer}
\affiliation{Institut f\"ur Theoretische Physik, Universit\"at
Regensburg, D-93040 Regensburg, Germany}
\author{Antonio Pineda}
\email{AntonioMiguel.Pineda@uab.es}
\affiliation{Grup de F\'{\i}sica Te\`orica, Universitat
Aut\`onoma de Barcelona, E-08193 Bellaterra, Barcelona, Spain}
\author{Christian Torrero}
\affiliation{
Dipartimento di Fisica e Scienze della Terra \&
Istituto Nazionale di Fisica Nucleare (gruppo collegato di Parma),
viale G. P. Usberti 7/A,
I-43124 - Parma, Italy}
\date{\today}

\begin{abstract}
\noindent
We determine the infinite volume coefficients of the perturbative expansions
of the self-energies of static sources in the fundamental and 
adjoint representations in $\mathrm{SU}(3)$
gluodynamics to order $\al^{20}$ in the strong coupling
parameter $\al$. We use numerical stochastic perturbation theory, where
we employ a new second order integrator and twisted boundary
conditions. The expansions are obtained in lattice regularization with the
Wilson action and two different discretizations of the 
covariant time derivative within the Polyakov loop. 
Overall, we obtain four different perturbative series.
For all of them the high order coefficients display the factorial growth 
predicted by the conjectured renormalon picture,
based on the operator product expansion. This enables us
to determine the normalization 
constants of the leading infrared renormalons of heavy quark
and heavy gluino pole masses 
and to translate these into the modified
minimal subtraction scheme ($\MS$). We also estimate
the four-loop $\beta$-function coefficient of the lattice scheme.
\end{abstract}
\pacs{12.38.Gc,11.15.Bt,12.39.Hg,12.38.Cy,12.38.Bx}
\maketitle

\section{Introduction}

Perturbative expansions in the coupling parameter
$\al$ of four dimensional non-Abelian
gauge theories, $\sum^{\infty}_{n=0}c_n\al^{n+1}$, are expected to be asymptotic. 
The structure of the operator product expansion (OPE)
determines one particular pattern of 
asymptotic divergence. This is usually
named a renormalon~\cite{Hooft} or, more specifically, an infrared renormalon. 
Its existence has not been proven. It could only be tested in QCD by assuming 
the dominance of $\beta_0$-terms, which amounts to an effective Abelianization 
of the theory, or in the two-dimensional $\mathrm{O}(N)$ model~\cite{David:1982qv}, where it is suppressed by powers of $1/N$. Moreover, the possible non-existence or irrelevance
of renormalons in Quantum 
Chromodynamics has been suggested in several papers, 
see, e.g.\ Refs.~\cite{Suslov:2005zi,Zakharov:2010tx} and references therein.
This has motivated dedicated 
high order perturbative expansions of the 
plaquette, see,
e.g. Refs.~\cite{DiRenzo:1995qc,Burgio:1997hc,Horsley:2001uy,Rakow:2005yn}, 
in lattice regularization, with conflicting conclusions.
Powers as high as $\al^{20}$
were achieved in the most recent simulation~\cite{Horsley:2012ra}.
However, the expected asymptotic behavior was not seen. 
A confirmation of this ``non-observation'' in the infinite volume
limit would
significantly affect phenomenological analyses
of data from high energy physics experiments where renormalon physics plays
a fundamental role. This is certainly so in heavy quark physics, where
addressing the pole mass renormalon is compulsory 
for almost any precise computation, such as for
determinations of the heavy quark masses in the
$\MS$ scheme, the decay of heavy hadrons,
or heavy quarkonium physics. 

Fortunately, in a recent letter, the existence of renormalons in 
quantum gluodynamics has been unambiguously established~\cite{Bauer:2011ws}.
The quantities studied were the self-energies of static sources in the
fundamental and adjoint representations.
This analysis clearly identified the reasons 
for the previous non-detection of the renormalon-associated
asymptotic behavior of the plaquette.
In lattice regularization with the Wilson action, renormalon dominance only sets
in at very high orders in $n$. 
In the case of the static self-energy, an operator of dimension
$d=1$, the renormalon behavior was confirmed for $n\gtrsim 9$. Therefore,
for the plaquette and the associated gluon condensate,
an operator of dimension four, we expect $n \sim 4\times 9$
to be necessary
to confirm the expected asymptotic behavior, an order that
is quite beyond those reached so far in
simulations. On top of this, it was 
shown that for presently reachable volumes the proper
incorporation of the leading finite size effects (FSE) is
required to obtain the correct infinite volume limit, something that had
not been done previously either. Finally, in
Ref.~\cite{Bauer:2011ws}
preliminary results for the normalization of the
renormalon were obtained, which turned 
out to be perfectly consistent with
expectations from continuum computations in the $\MS$ scheme. 
In this article we provide greater detail on these simulations
and our analysis methods, present finalized
results, and further extend this previous study.

This article is organized as follows. In Sec.~\ref{sec:NSPT}
we review numerical stochastic perturbation theory (NSPT),
our improvements on previously existing techniques,
and the specific aspects of the lattice computation relevant for our case.
In Sec.~\ref{Sec:deltam} we define our primary observable:
the self-energy of a static source, and detail the expected
asymptotic behavior of its perturbative 
expansion due to the leading renormalon. In
Sec.~\ref{Sec:Polya} we define the Polyakov loop, 
relate this to the static
self-energy, and 
explain how our primary data sets are obtained. 
In Sec.~\ref{sec:fse1} we present a theoretical study of
the leading FSE
and how these will
affect the signatures of renormalon dominance.
Subsequently, in Sec.~\ref{sec:FiniteNT},
we investigate, mostly numerically, subleading FSE
that may pollute our data and estimate their systematics.
In Sec.~\ref{Sec:fits} we determine the infinite volume
coefficients of the perturbative expansion, study their renormalon structure, and extract universal results
in the lattice and $\MS$ schemes, before we conclude.

\section{Lattice implementation}
\label{sec:NSPT}
Below we discuss the simulation method and its implementation.
After a brief introduction into NSPT 
we detail a new second order integrator, introduce
twisted boundary conditions and link smearing.

\subsection{Stochastic Quantization and NSPT}
Stochastic Quantization (SQ)~\cite{PW} enables the calculation
of expectation values in quantum field theories and 
presents an alternative to, for instance, the path integral formalism. 
In recent years, SQ was employed in several studies
within different fields of physics,
ranging from the quark-gluon plasma~\cite{Svet},
even addressing the notorious sign problem
of QCD at non-vanishing baryon densities~\cite{Aarts,Cristoforetti:2012su}, 
to quantum
gravity~\cite{Loll}.
SQ turns out to be efficient also from the point of view of computer
simulations
due to the absence of any global accept/reject step
thus allowing, in principle, for a fast update of the system under
consideration. 
The draw-back is the requirement to span
a range of integration step
sizes, to enable an extrapolation to continuous stochastic time.

For simplicity, we
assume a scalar field $\phi(x)$ depending on spacetime $x$ and
dynamics governed by an action $S[\phi]$. 
The core of SQ, the Langevin equation, then reads
\begin{equation}
\label{eq:Langevin}
\frac{\partial\phi(x,t)}{\partial t} = -\frac{\partial S[\phi]}{\partial\phi(x,t)} - \eta(x,t)\, ,
\end{equation}
where $t$ is the so-called stochastic time.
The $\eta(x,t)$ is a Gaussian noise variable with the properties
\begin{align}
\langle \eta(x,t)\rangle_{\eta} &= 0 \, , \nonumber\\
\langle \eta(x,t)\eta(x',t')\rangle_{\eta}&= 2\delta(x-x')\delta(t-t')\,. 
\end{align} 
The subscript ``$\eta$" stands for an average over the noise.
Given a generic observable $A(\phi)$, it can be shown\footnote{
For a proof in perturbation theory, see Ref.~\cite{Floratos:1982xj}.} 
that the time average
\begin{equation}
\overline{A(\phi)} = \lim_{T\rightarrow+\infty}\frac{1}{T}\int_0^T\!\!dt\,
A(\phi)
\end{equation}
coincides with the expectation value on the quantum vacuum, i.e.,
\begin{equation}
\label{eq:MeanSQ}
\overline{A(\phi)} = \frac{1}{Z}\int\! [d\phi]\, A(\phi)\, e^{-S[\phi]}\,,
\end{equation}
where $Z$ is the partition function. 

If the degrees of freedom of the system under 
consideration are not scalar but obey a group structure, as it is the 
case for lattice QCD, the above machinery has to be 
modified accordingly (numerical stochastic 
perturbation theory, NSPT~\cite{DRMMOLatt94,DRMMO94}, for a review 
see Ref.~\cite{DR0}).
In lattice simulations, spacetime is
discretized by introducing a four-dimensional hyper-cube of
$N_S^3\times N_T$ sites, 
where asymmetric volumes $N_S\neq N_T$ are legitimate. 
A peculiarity of NSPT is that no mass gap can be generated in
perturbation theory. Hence the lattice spacing $a$
is neither set nor determined a posteriori, 
so any NSPT-related reference to $a$ is purely formal.
For instance, the limit $N_S\rightarrow\infty$ can either be
interpreted as the infinite volume limit $L=aN_S\rightarrow\infty$
at fixed $a$ or as the continuum limit $a\rightarrow 0$ at fixed
lattice extent $L$ in physical units.
Lattice sites $n$ are referenced by their spatial and temporal
coordinates,
$n_i\in\{0,\ldots,N_S-1\}$ and $n_4\in\{0,\ldots,N_T-1\}$, respectively.

The gauge degrees of freedom 
$A^R_{\mu}(x)$ in the continuum are elements of the Lie Algebra of $\mathrm{SU}(3)$
in representation\footnote{Representation $R$ has the dimension
$d_R$. Here we consider two representations:
the fundamental triplet ($d_R=3$) and the adjoint octet
($d_R=8$).} $R$. 
On the lattice these are implemented as compact
link variables $U^R_{\mu}(n)\approx 
e^{iA^R_{\mu}[(n+1/2)a]}\in\mathrm{SU(3)}$,  connecting the sites $n$ and 
$n+\hat{\mu}$, where $\hat{\mu}$ denotes a unit vector in direction
$\mu$. 

The straightforward generalization of the
Langevin equation \Eqre{eq:Langevin} to fundamental link variables
reads
\begin{equation}
\label{eq:LangevinU}
\frac{\partial}{\partial t}U_{\mu}(n,t)=-i\sum_{a}T^a\big[\nabla^a_{\!\!n,~\!\!\mu}S[U]+\eta^a_{\mu}(n,t)\big]U_{\mu}(n,t)\, ,
\end{equation}  
where $S[U]$ is the gauge action and $T^a, a=1,\dots,8$ are the
traceless Hermitian generators of the $\mathrm{SU}(3)$ Lie algebra with the
normalization $\mathrm{tr}(T^aT^b)=\frac{1}{2}\delta_{ab}$. 
We define
the derivative within \Eqre{eq:LangevinU} of a function $f(U)$
with respect to a Lie group variable $U$ following Ref.~\cite{Drum}:
\begin{equation}
\label{eq:nabla}
f\left(e^{i\sum_aT^a\omega^a}U\right) = f(U) + \sum_a\omega^a\nabla^af(U)
+ \mathcal{O}(\omega^2)\,,
\end{equation}
where $\omega^a$ are small real parameters.

Perturbative lattice simulations up to $\nmax$
loops become possible by a \emph{formal} weak coupling expansion
of the gauge fields. In the algebra and group this reads
\begin{align}
\label{eq:TaylorU}
 A &= A^{(1)}\beta^{-\frac{1}{2}}  + A^{(2)}\beta^{-1}  + \dots+A^{(2\nmax)}\beta^{-\nmax}\,,\\
 U &= \mathds{1}+U^{(1)}\beta^{-\frac{1}{2}}  + U^{(2)}\beta^{-1}  + \dots+U^{(2\nmax)}\beta^{-\nmax}\,.\nonumber
\end{align}
Above, $\beta$ denotes
the lattice coupling and relates to the strong coupling parameter
as $\beta^{-1}=g^2/6=(2\pi/3)\alpha$. 
Note that while the $A^{(i)}$
belong to the Lie algebra of $\mathrm{SU}(3)$,
the $U^{(i)}$ are no group elements.
$U$ however is, up to terms of $\mathcal{O}(\beta^{-(\nmax+1)})$,
an $\mathrm{SU}(3)$ group element. By
Taylor expanding the exponent and logarithm of the two series,
respectively, 
one can conveniently switch between algebra and group representations.
Plugging the expansion \Eqre{eq:TaylorU} into a discretized
version of the stochastic differential equation~\Eqre{eq:LangevinU}, one
finds that the noise directly
acts only on $U^{(1)}$ while the evolution of higher orders is
governed
by a hierarchical system of ordinary differential equations. 
In particular, the evolution of a given order $U^{(i)}$ in
stochastic time only depends on
preceding orders $1,\dots,i-1$ so that
a truncation at finite $\nmax$ is possible. 

The naive computational effort of NSPT scales like $n^2_{\mathrm{max}}$
and the memory requirement like $\nmax$, compared to
a factorial growth of the number of diagrams
$\sim \nmax !$ in conventional perturbation theory.
This makes high order expansions feasible.
On an absolute scale, computation time of course becomes an issue for large lattice volumes or high $\nmax$, 
requiring optimizations of the NSPT algorithm.
This study would have exceeded our present computer resources
had we not used 
an improved numerical
algorithm to evolve the Langevin equation~\Eqre{eq:LangevinU}.
Its advantages were detailed in Ref.~\cite{Torrero:08}. 
Below we present the algorithm in detail.

\subsection{The second-order integration scheme}
\label{sec:secondorder}
The numerical integration of the Langevin equation \Eqre{eq:LangevinU}
requires the discretization of the stochastic time $t$, introducing
a time step $\epsilon$ 
($t=t_m=m\epsilon$ with integer $m$) 
and a prescription for the $t$-derivative in \Eqre{eq:LangevinU}. 
Revisiting the scalar example,
schematically the updating step for the $i$th degree of
freedom $\phi_i$ reads\footnote{We have replaced
the dependence on discretized spacetime coordinates $n$
by an index $i$ for simplicity.}
\begin{equation}
\phi_i^{(m+1)} = \phi_i^{(m)} - f_i^{(m)}\, ,
\end{equation}  
where the bracketed superscript labels the evolution in Langevin time 
$t=m\epsilon$
and $f_i$ is a force term. 
In the simplest (Euler) integration scheme the force is given by
\begin{equation}
\label{eq:EulerForcePhi}
f_i^{(m)} = \epsilon\nabla_{\!\!i}S^{(m)} + \sqrt{\epsilon}\,\eta_i^{(m)}\, ,
\end{equation}
with the functional derivative $\nabla$ defined in \Eqre{eq:nabla}
for gauge theories and
$\eta_i^{(m)}\! =\! \sqrt{\epsilon}~\!\eta(n,t=m\epsilon)$. 

Information on how the discretization changes the equilibrium
distribution relative to the continuous-time expression
of \Eqre{eq:MeanSQ} 
can be drawn from the Fokker-Planck equation. 
To work this out, we label the probability distribution
after $m+1$ updates as $\mathcal{P}^{(m+1)}(\phi)$: 
by defining $W(\phi'\leftarrow\phi)$ as the probability 
of jumping from configuration $\phi$ to configuration $\phi'$, 
we obtain the equality
\begin{align}
\mathcal{P}^{(m+1)}(\phi') &=
 \int\![d\phi]\,W(\phi'\leftarrow\phi)\mathcal{P}^{(m)}(\phi)=\nonumber\\
&= \int\![d\phi][d\eta]\,\prod_i\delta(\phi_i'-\phi_i+f_i)\,\mathcal{P}^{(m)}(\phi)\,. 
\end{align} 
The above product extends over all degrees of freedom 
and we have rewritten the probability of moving from $\phi$ to $\phi'$ 
in terms of $\delta$-functions, involving the noise (that is implicit in $f_i$). 
After some algebra,\footnote{
Essentially, one represents each $\delta$-function as a
Fourier integral,
Taylor-expands in the force term, expresses each
power of the expansion by suitable derivatives 
with respect to the $\phi$s and integrates by parts.}
one obtains
\begin{equation}
\label{eq:Pn+1}
\mathcal{P}^{(m+1)}(\phi) = \mathcal{P}^{(m)}(\phi) + \sum_{j=1}^{\infty}\frac{1}{j!}\nabla_{i_1}\ldots\nabla_{i_j}\left[\langle f_{i_1}\ldots f_{i_j}\rangle_{\eta}\mathcal{P}(\phi)\right]\,.
\end{equation}
We recall that $\mathcal{P}^{(m+1)}(\phi)=\mathcal{P}^{(m)}(\phi)$
at equilibrium, 
insert force terms into \Eqre{eq:Pn+1} and expand with respect to $\epsilon$. 
This leads to the identity
\begin{equation}
0 = \nabla_{\!\!i}\left\{\nabla_{\!\!i} \overline{S}[\phi] + \nabla_{\!\!i}\right\}\mathcal{P}(\phi)\ ,
\end{equation}
whose solution reads $\mathcal{P}(\phi)\propto e^{-\overline{S}[\phi]}$ with
\begin{equation}
\label{eq:Sepsilon}
\overline S[\phi] = S[\phi] + \epsilon S_1[\phi] + \epsilon^2 S_2[\phi] + \ldots\, .
\end{equation}
Within the above equation, $S[\phi]$ is the original action of
\Eqre{eq:Langevin}. 
Thus, the correct equilibrium distribution and, consequently, 
\Eqre{eq:LangevinU} is recovered in the limit
$\epsilon\rightarrow 0$.
In the Euler scheme, for example, $\overline S[\phi]$ is given by
\begin{equation}
\label{eq:EquilibEulerPhi}
\overline S[\phi] = S[\phi] + \frac{\epsilon}{4}\sum_i\Big(2\nabla_{\!\!i}\nabla_{\!\!i} S[\phi] - \nabla_{\!\!i} S[\phi]\nabla_{\!\!i} S[\phi]\Big) + \mathcal{O}(\epsilon^2)\ .
\end{equation}

We detailed the formalism for a scalar field $\phi$. 
In the case of non-Abelian $\mathrm{SU}(N_c)$ gauge theory,
the discretized Langevin update reads\footnote{
The index $i$ now contains both spacetime position $n$ and direction $\mu$.}  
\begin{equation}
U^{(m+1)}_i = e^{-i\sum_a\!T^a\!f^a_i}~\!U^{(m)}_i\, ,
\end{equation}
where the force term in the Euler scheme is given by the analogue of
\Eqre{eq:EulerForcePhi}:
\begin{equation}
\label{eq:EulerForceU}
f_i^a = \epsilon\nabla^a_{\!\!i}S[U] + \sqrt{\epsilon}\,\eta_i^a\, .
\end{equation}
With the group derivative defined as in \Eqre{eq:nabla}
the above procedure can be repeated for non-Abelian degrees of freedom, 
again leading to a Fokker-Planck equation. 
The only difference lies in the fact that group derivatives do not commute. 
More precisely, in the continuum
\begin{equation}
[\nabla^a_{\!\!x,\mu},\nabla^b_{\!\!y,\nu}] = -f^{abc}\ \!\nabla^c \delta_{xy}\delta_{\mu\nu}\, ,
\end{equation}
where $f^{abc}$ are the structure constants of the Lie algebra. 
Obviously, this has a non-trivial impact on the equilibrium distribution $\overline{S}[U]$ at $\epsilon>0$. 
For instance, plugging \Eqre{eq:EulerForceU} into the Fokker-Planck equation,
we obtain
\begin{equation}
\label{eq:EquilibEulerU}
\overline S[U] = \left(1+\frac{\epsilon C_A}{4}\right)S[U] + \frac{\epsilon}{4}\sum_{i,a}\left(2\nabla^a_{\!\!i}\nabla^a_{\!\!i} S[U] - \nabla^a_{\!\!i} S[U]\nabla^a_{\!\!i} S[U]\right) + \mathcal{O}(\epsilon^2)\ ,
\end{equation}
where $C_A=N_c$ is the quadratic Casimir invariant of the adjoint representation
of $\mathrm{SU}(N_c)$.

From Eqs.~\eqref{eq:EquilibEulerPhi} and \eqref{eq:EquilibEulerU} it
is evident that 
numerical simulations with different values of $\epsilon$ 
are necessary to extrapolate to continuous stochastic time
 $\epsilon\to 0$ and to recover \Eqre{eq:LangevinU} and the
continuum distribution. 
Simulations at small $\epsilon$ obviously are more costly
and it is tempting to keep $\epsilon$ as large as possible. 
However, for large time steps corrections to the leading linear
dependence will become sizable and extrapolations to $\epsilon=0$
less controlled.

A reduction in computer time while maintaining a safe
$\epsilon\rightarrow 0$ extrapolation becomes
possible by employing higher-order integration schemes.
To our knowledge, Runge-Kutta schemes exist up to the
third order for Abelian theories~\cite{Helf,Gren} 
(the general solution to the Fokker-Planck equation is known), 
and up to the second order in $\epsilon$ 
for non-Abelian $\mathrm{SU}(N_c)$ theories~\cite{Bat,Fuk}. 
In the latter case, only one variant of the general solution
is published, 
namely the two-step algorithm
\begin{align}
\label{eq:AnsatzRunge1}
U'_i&= e^{-i\sum_a\!T^a\!\left(\epsilon\nabla^a_{\!\!i}S[U] + \sqrt{\epsilon}\eta_i^a\right)}\,U^{(m)}_i \, ,\\
\label{eq:AnsatzRunge2}
U^{(m+1)}_i &= e^{-i\sum_a\!T^a\!\left(\frac{1}{2}\epsilon\nabla^a_{\!\!i}S[U] + \frac{1}{2}\epsilon\nabla^a_{\!\!i}S[U'] + \frac{C_{\!A}}{6}\epsilon^{\!2}\nabla^a_{\!\!i}S[U'] + \sqrt{\epsilon}\eta_i^a\right)}\,U^{(m)}_i \, ,
\end{align}
where $S[U]$ and $S[U']$ stand for the action computed using the fields $U^{(m)}$ and $U'$, respectively.
We refer to this second-order integrator as the
``BF scheme''~\cite{Bat,Fuk}.
Note that the evolution cannot be factorized into sweeps
involving single link updates:
both $U_i'$ and $U_i^{(m+1)}$
have to be computed for all links $i$, prior to
the replacement of the
original field $U_i^{(m)}$. In particular, in the second
step both $S[U]$ and $S[U']$ are needed. This requires three
copies to be kept in memory concurrently of
$2\nmax+1$ orders of complex three by three  matrices for each lattice link.

Below we derive the general solution 
and
provide an optimized alternative to
Eqs.~\eqref{eq:AnsatzRunge1} and \eqref{eq:AnsatzRunge2} 
which not only saves matrix additions but also reduces the memory requirements.
The general ansatz for the second-order algorithm reads
\begin{align}
\label{eq:Ansatz1}
U'_i&=e^{i\sum_a\!T^a\!\left(k_1\epsilon\nabla^a_{\!\!i}S[U] + k_2\sqrt{\epsilon}\eta_i^a\right)}\,U^{(m)}_i \, ,\\
\label{eq:Ansatz2}
U^{(m+1)}_i&= e^{-i\sum_a\!T^a\!\left(k_3\epsilon\nabla^a_{\!\!i}S[U] + k_4\epsilon\nabla^a_{\!\!i}S[U'] + k_5C_A\epsilon^{\!2}\nabla^a_{\!\!i}S[U'] + k_6\sqrt{\epsilon}\eta_i^a\right)}\,U^{(m)}_i \,.
\end{align}
Plugging the force term of \Eqre{eq:Ansatz2} 
into the Fokker-Planck equation and Taylor-expanding the derivative of
$S[U']$, after some algebra some constraints
are obtained: at $\mathcal{O}(\epsilon^0)$ 
the non-Abelian analogue of \Eqre{eq:Sepsilon} yields
\begin{equation}
k_3=1-k_4\,,\quad k_6^2=1\ ,
\end{equation}
in order to recover the correct $\epsilon\rightarrow 0$ distribution,
while the elimination of terms proportional to $\epsilon$ (using
$k_3=1-k_4$) results in
\begin{align}
k_1 &= \frac{1-4k_4\pm 2\sqrt{2k_4(2k_4-1)}}{2k_4}\, ,\\
k_2 &= \frac{-2k_4\pm \sqrt{2k_4(2k_4-1)}}{2k_4k_6}\, ,\\
k_5 &= \frac{-1+6k_4\mp3\sqrt{2k_4(2k_4-1)}}{12}\,,
\end{align}
where $k_4$ and $k_6=\pm 1$ can be chosen freely.
The BF scheme
is recovered setting $k_4=\frac{1}{2}=k_3$, $k_6=1$.
The choice $k_4=k_6=1$, however, further simplifies the algorithm: 
\begin{align}
\label{eq:SolutionRunge1}
U'_i&= e^{i\sum_a\!T^a\!\left(\frac{-3+2\sqrt{2}}{2}\epsilon\nabla^a_{\!\!i}S[U] - \frac{2-\sqrt{2}}{2}\sqrt{\epsilon}\eta_i^a\right)}\,U^{(m)}_i \, ,\\
\label{eq:SolutionRunge2}
U^{(m+1)}_i &= e^{-i\sum_a\!T^a\!\left(\epsilon\nabla^a_{\!\!i}S[U'] + \frac{(5-3\sqrt{2})C_{\!A}}{12}\epsilon^{\!2}\nabla^a_{\!\!i}S[U'] + \sqrt{\epsilon}\eta_i^a\right)}\,U^{(m)}_i\,.
\end{align}
The gain of this variant is twofold: besides saving a
matrix addition when computing the force term 
of \Eqre{eq:SolutionRunge2} instead of \Eqre{eq:AnsatzRunge2}, 
there is no need to store (or to
recompute) $\nabla_{\!\!i}S[U]$. After the intermediate step
the original fields (and/or $S_i[U]$) can be overwritten, reducing
the memory requirement by one third.
 
We tested the integrator defined through
Eqs.~\eqref{eq:SolutionRunge1} and \eqref{eq:SolutionRunge2} within NSPT: 
due to the need to rescale the time step $\epsilon\mapsto \epsilon/\beta$
(for details see, e.g., Ref.~\cite{DR0}), after inserting the
perturbative expansion into the discretized Langevin equation,
it turns out that the contribution proportional to $C_A$ in the
force term of \Eqre{eq:SolutionRunge2} only affects
the two-loop level and beyond. 

\begin{table}
\caption{\it Comparison on $N^4$ lattices between analytical
and NSPT results for the one-loop coefficient $p_0$ of the plaquette: 
The $\epsilon$-values used in the extrapolations range from 0.04 to 0.07.
The analytical result is given by $2(1-N^{-4})$.}
      \label{tab:1loop}
\begin{ruledtabular}
\begin{tabular}{c|cccc}
$N$ & analytical &Euler&2nd-order BF& new 2nd-order\\ 
        \hline
        4 &  1.9921875  & 1.9931 (6) & 1.9922 (9) & 1.9924 (7) \\
        6 &  1.9984568  & 1.9985 (3) & 1.9987 (3) & 1.9986 (3) \\
        8 &  1.9995117  & 1.9997 (2) & 1.9992 (3) & 1.9996 (3) \\
        10 & 1.9998000  & 1.9996 (1) & 2.0001 (2) & 2.0001 (2) \\
        12 & 1.9999035  & 1.9998 (1) & 1.9999 (1) & 1.9998 (1)
\end{tabular}
\end{ruledtabular}
\end{table}
\begin{table}
\caption{\it Comparison on $N^4$ lattices between diagrammatic
lattice perturbation theory (DLPT) and NSPT results for the two-loop coefficient $p_1$ of the plaquette.}
      \label{tab:2loop}
\begin{ruledtabular}
\begin{tabular}{c|cccc}
$N$ & DLPT &Euler&2nd-order BF& new 2nd-order\\ 
        \hline
        4 &  1.20370366  & 1.2020(15) & 1.2005(17) & 1.2012(17) \\
        6 &  1.21730787  & 1.2173 (7) & 1.2166 (8) & 1.2180 (8) \\
        8 &  1.21965482  & 1.2203 (4) & 1.2178 (7) & 1.2199 (6) \\
        10 & 1.22031414  & 1.2203 (3) & 1.2204 (4) & 1.2212 (6) \\
        12 & 1.22055751  & 1.2208 (2) & 1.2200 (3) & 1.2204 (2)
\end{tabular}
\end{ruledtabular}
\end{table}

In Tables~\ref{tab:1loop} and \ref{tab:2loop}
we compare one- and two-loop plaquette coefficients
$p_0$ and $p_1$, defined through
\begin{equation}
\langle U_{\Box}\rangle = 1-p_0\beta^{-1}-p_1\beta^{-2}-\ldots\,.
\end{equation}
These were computed using the new
second-order algorithm,
diagrammatic lattice perturbation theory,
the Euler integrator and the BF scheme for
different symmetric volumes $N^4$ with periodic boundary
conditions (PBC).
For the Euler integrator, the fit function employed
in the extrapolation is constant plus linear
while for the second-order schemes the ansatz
is constant plus quadratic
(in the latter case, we checked that the coefficients of
terms linear in $\epsilon$ indeed vanish within errors 
when using a linear plus quadratic fit function). 
In all the cases we find agreement between the methods
within two standard deviations.

\begin{figure}[h]
	\centerline{\includegraphics[width=.9\textwidth,clip]{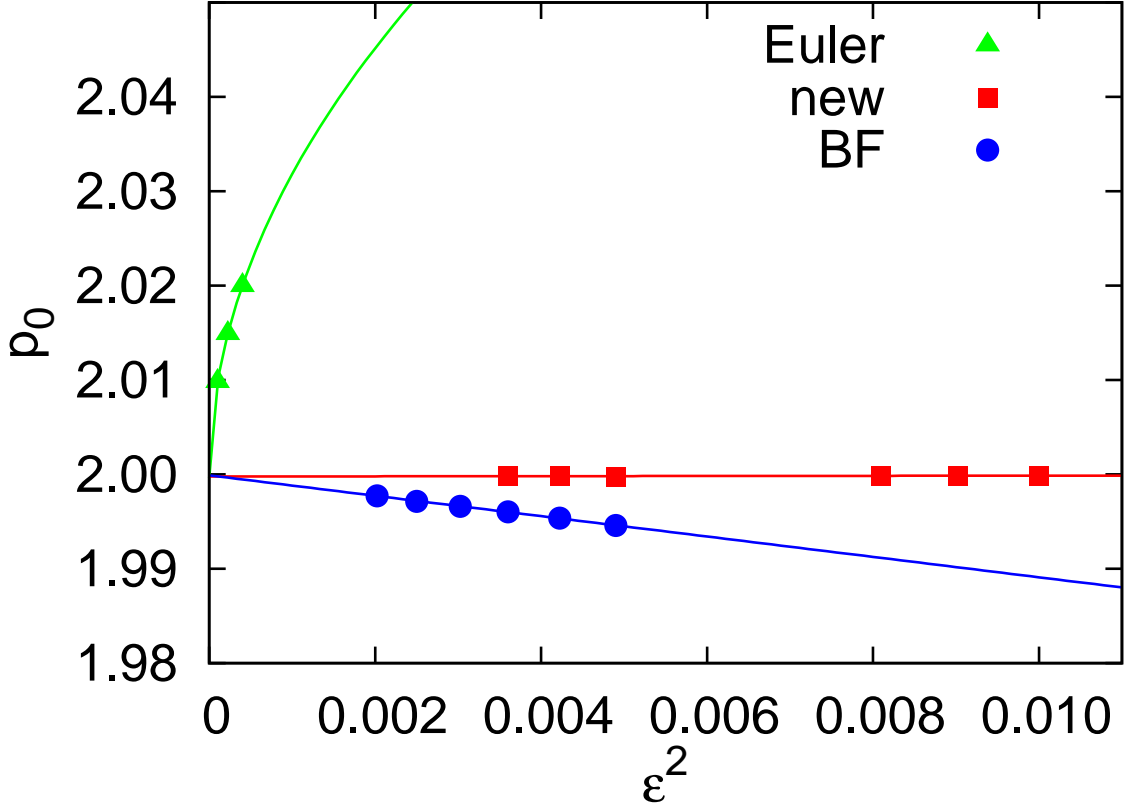}}
      \caption{\it The one-loop plaquette coefficient $p_0$ vs.\ $\epsilon^2$ for $N^4=10^4$:
Euler integrator, BF scheme and the new second order integrator.
      \label{fig:1loop}}
\end{figure}
\begin{figure}[h]
    \centerline{\includegraphics[width=.9\textwidth,clip]{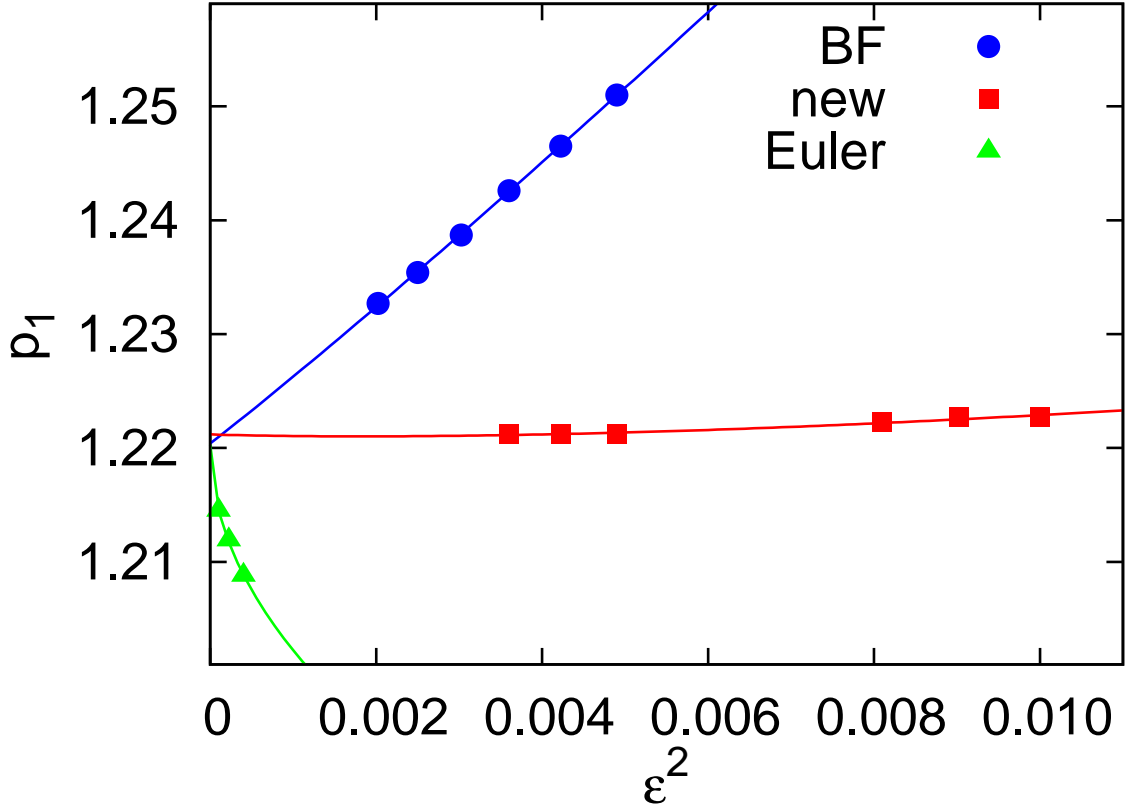}}
      \caption{\it The two-loop plaquette coefficient $p_1$ vs. $\epsilon^2$ for $N^4=10^4$:
Euler integrator, BF scheme and the new second order integrator.
      \label{fig:2loop}}
\end{figure}

Figures~\ref{fig:1loop} and \ref{fig:2loop} illustrate
the finite-$\epsilon$ plaquette results for the three integrators: 
while all sets extrapolate to the same limit within error bars, 
the ones corresponding to the new second-order scheme are clearly
much flatter than the others. In particular,
the $\epsilon^2$-dependence of the new second-order integrator
is greatly reduced compared to the BF scheme
of Eqs.~\eqref{eq:AnsatzRunge1} and \eqref{eq:AnsatzRunge2}.
Note that we allowed for cubic terms in the curves drawn
for the second order integrators.
We will see in Sec.~\ref{Sec:Polya} below
that for the observables of interest in
this work extrapolations in $\epsilon^2$ are so flat that
in most cases a non-trivial slope cannot be resolved
within statistical errors.

\subsection{Twisted Boundary Conditions\label{sec:Twist}}
Instead of PBC, one can also
impose twisted boundary conditions
(TBC)~\cite{tHooft79,Parisi83,Luscher86,Arroyo88}
\begin{align}
   \label{eq:TBCU}
   \Umu(n+N_S\hat\nu) &= \Omega_\nu \Umu(n) \Omega_\nu^\dagger\,,\\
\Umu(n-N_S\hat\nu) &= \Omega_\nu^\dagger \Umu(n) \Omega_\nu\,,
\end{align}
where the links that pierce a lattice boundary of a twisted (spatial) direction~$\hat\nu$ are multiplied by so-called twist matrices~$\Omega_\nu$  that must satisfy
\begin{align}
   \Omega_\mu\Omega_\nu &= z \, \Omega_\nu\Omega_\mu,\label{eq:Omega}\\
   \Omega_\nu^3 &= \left(-1\right)^{N-1} \mathds{1}\,.
\end{align}
Here $z \in \{\mathds{1},e^{i2\pi/3}\mathds{1},e^{i4\pi/3}\mathds{1}\}$
is an element of the center of $\mathrm{SU}(3)$.
The
condition~\Eqre{eq:Omega} guarantees that the value of
the transported link $\Umu\left(n+N_S\hmu+N_S\hnu\right)$ 
is independent of the order with which two twisted boundaries~$\mu,\nu$
are transversed.  
Gauge transformations $\Lambda(n)$, which rotate the link variables according to
\begin{align}
\label{eq:Rotate}
\Umu(n) \mapsto \Lambda(n)\Umu(n)\Lambda^{\dagger}(n+\hmu)
\end{align}
must obey the same TBC Eq.~\eqref{eq:TBCU}. 

The measure
as well as Wilson loops without net winding numbers
across boundaries (such as the elementary plaquette
within the action)
are invariant under the transformation
\begin{equation}
\label{eq:Z3}
   \Umu(n) \to z \Umu(n)\,,
   \quad \forall\, n \in \{n:n \cdot \hat\mu = \mbox{const.}\}\,.
\end{equation}
TBC rely on this center symmetry of the $\mathrm{SU}(3)$
gauge action and measure,
and can be implemented for the link update either
by multiplying the plaquettes in corners of twisted hyper-planes 
with suitable center elements, 
or by imposing~\Eqre{eq:TBCU} with an explicit choice of $\Omega_\nu$.
We implemented the latter using
\begin{equation}
   \label{eq:ExplicitOmega}
   \Omega_{1} = \left( 
   \begin{array}{ccc}
   0 & 1 & 0 \\
   0 & 0 & 1 \\
   1 & 0 & 0
   \end{array} \right)\,,\quad
   \Omega_{2} = \left( 
   \begin{array}{ccc}
   \zeta^* & 0 & 0 \\
   0 & 1 & 0 \\
   0 & 0 & \zeta
   \end{array} \right)\,,\quad
  \Omega_{3} = \Omega_{2}\Omega_{1}^{2} = \left( 
   \begin{array}{ccc}
   0 & 0 & \zeta^* \\
   1 & 0 & 0 \\
   0 & \zeta & 0
   \end{array} \right)\,,
\end{equation}
where $\zeta=e^{2i\pi/3}$, $\zeta^*=1/\zeta$.
This choice is arbitrary up to global unitary transformations: 
as long as \Eqre{eq:Omega} is satisfied, 
the resulting physical amplitudes will not depend on the explicit choice of~$\Omega_\nu$. 
As the subscripts indicate, we impose the twist for all spatial directions.
Twists in two directions have a non-trivial effect too,
while twisting only one direction can be absorbed into a re-definition
of the link variables. The effect of twist is twofold:
TBC eliminate zero modes which otherwise require an explicit
subtraction~\cite{DR0}. Furthermore, at least at low orders in
perturbation theory, TBC reduce finite size effects as the possible
gluon momenta  are restricted to integer multiples~\cite{Luscher86} of
   \begin{equation}
\label{eq:kQuantize}
   p_\nu = \left\{ \begin{array}{cl}
   \frac{2\pi}{3N_{\nu}}\,, & \nu = \mbox{twisted direction}\,, \\
   \frac{2\pi}{N_{\nu}}\,, & \nu = \mbox{periodic direction}\,.
   \end{array} \right.
\end{equation}
This means that gluon momenta in twisted spatial directions reach values as
low as ${2\pi}/{(3N_S)}$, compared to ${2\pi}/{N_S}$
in periodic directions. 
So, roughly speaking, the modes in a twisted direction behave as if
the corresponding lattice extent was $3N_S$ instead of $N_S$. 
We refer to the cases of twists applied to two and three
directions as TBCxy and TBCxyz, respectively.

\begin{table}
\caption{\it Plaquette coefficients.
$4^4$ PBC: DLPT (first two orders) and NSPT (remaining orders~\protect\cite{Perlt:privat}).
$4^4$ TBCxy and TBCxyz: DLPT (first order) and NSPT (remaining orders).
$32^4$ PBC: DLPT (first two orders) and NSPT ($\mathcal{O}(\beta^{-3})$~\protect\cite{DR0}). 
Infinite volume: DLPT from Ref.~\protect\cite{Alles:1998is}, using
the lattice integrals of Ref.~\protect\cite{Luscher:1995np}.
For all NSPT data, the $\epsilon\rightarrow 0$
extrapolation was carried out.\label{tab:plaq}}
\label{tab:PBCTBC}
\begin{ruledtabular}
\begin{tabular}{c|ccccc}
order&$4^4$ PBC&$4^4$ TBCxy&$4^4$ TBCxyz&$32^4$ PBC&$\infty^4$\\\hline
$\beta^{-1}$&1.9921875& 2 & 2&1.9999981&2\\
$\beta^{-2}$&1.2037037& 1.2184(5)& 1.2200(3)&1.2207904&1.2279575\\
$\beta^{-3}$&2.887(3)& 2.955(2) & 2.957(2)&2.957(3)&2.9605(1)\\
$\beta^{-4}$&-9.05(1)& -9.41(1) & -9.40(1)&&\\
$\beta^{-5}$&-32.49(6)& -34.51(9) & -34.34(5)&&
\end{tabular}
\end{ruledtabular}
\end{table}
The effect of TBC is noticeable in particular on small lattice volumes,
as Table~\ref{tab:PBCTBC} illustrates for the average plaquette.
The two- and three-loop TBCxy and TBCxyz data obtained on
$4^4$~volumes are close to the infinite volume (as well as to
$32^4$ PBC) results at two and three loops. 
This clearly is not the case for $4^4$ PBC data.
Note that both analytical one-loop TBC coefficients
happen to be volume-independent on symmetric
lattices, due to cancellations between different
plaquette orientations.

\begin{figure}
\centerline{
\includegraphics[width=.49\textwidth,clip]{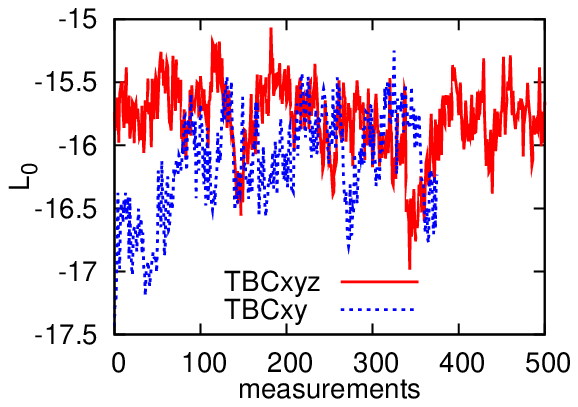}~
\includegraphics[width=.49\textwidth,clip]{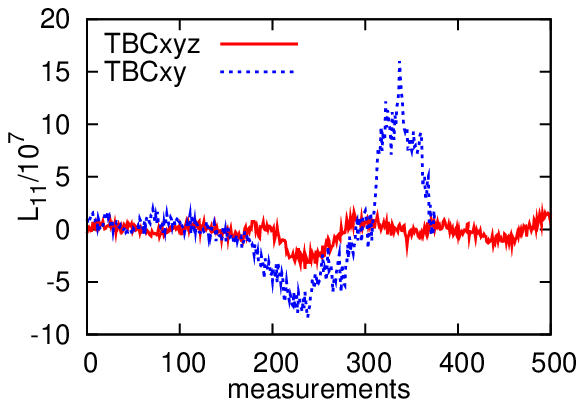}}
\caption{\it Stochastic time series of one-loop ($L_0$, left) and
12-loop ($L_{11}$, right) Polyakov loop coefficients for
TBCxyz (blue) and TBCxy (red) on $16^4$ lattices for
$\epsilon=0.05$.\label{fig:L1L12}}
\end{figure}

The situation is different for the Polyakov loop $L$
defined in \Eqre{eq:defpol} below. 
First of all, for this observable it matters whether it is obtained
in an untwisted or a twisted direction. 
We calculate $L$ in untwisted directions, 
for which no modification is necessary with respect to PBC,
and extract the static energy $\delta m$ via
Eqs.~\eqref{eq:defP}--\eqref{eq:deltam}. 
As it was shown in Ref.~\cite{Nobes:2001tf} for this observable, 
TBC significantly reduce FSE, resulting in a
much flatter extrapolation towards infinite volume. 
If this flatness at low orders was taken as the only criterion,
TBCxy would be the boundary condition of choice. 
However, it turns out that TBCxy has a draw-back compared to TBCxyz: 
in non-perturbative simulations only the latter prevents
tunneling between different Z(3) phases 
while TBCxy merely leads to a reduction compared to
PBC~\cite{Trottier:2001vj}. 
As a consequence, small volume TBCxy simulations were found to
fluctuate more and to return noisier signals than their TBCxyz counterparts. 
Regarding the statistical fluctuations
we made a similar observation, even though tunneling
between Z(3) sectors is not an issue in our NSPT simulations
since $U^{(0)}=\mathds{1}$. 
Fig.~\ref{fig:L1L12} shows stochastic time histories
obtained on $16^4$ volumes at fixed $\epsilon=0.05$ for TBCxy and TBCxyz
of the one-loop and 12-loop coefficients of the Polyakov loop.
While the trajectories of the one-loop coefficient~$L_0$
show a similar behavior for TBCxy as for TBCxyz, we observe
a peak in the twelve loop $L_{11}$ TBCxy measurement history,
which is symptomatic for TBCxy simulations.     
The enhanced numerical stability and smaller
fluctuations, in particular at large orders of
expansions,  motivated us to choose TBCxyz for this work. 
A better understanding of the origin of these differences between
TBCxy and TBCxyz would be desirable.

\subsection{Link smearing}\label{sec:smearing}
The lattice discretization of observables and action
is not unique. For instance one can construct Wilson loops
and Polyakov loops, replacing the link
variables $U_{\mu}(n)$ by fat or ``smeared'' links.
In the context of a lattice determination of
static potentials and of static-light meson masses this was for instance
done in Ref.~\cite{Bali:2005fu}, to reduce the self-energy, enabling
an improved signal to noise ratio at large Euclidean times.
As long as the smearing is an ultra-local procedure,
defined on the scale of a few lattice spacings,
making this replacement in a Polyakov loop corresponds to a
different choice of discretization of the static action.
Smearing is sometimes also used within the definition of
fermionic actions, see, e.g., Refs.~\cite{Blum:1996uf,Zanotti:2001yb,Capitani:2006ni}.

Several smearing methods are available, one of which is known as 
analytic or ``stout'' smearing~\cite{Morningstar:2003gk}. 
Stout links are automatically elements of the $\mathrm{SU}(3)$ group,
without a
numerically delicate projection into the group.
Therefore, implementing stout smearing within 
a perturbative expansion is straightforward. Stout smeared links
are obtained by the replacement 
\begin{equation} 
 \Umu(n)\mapsto U_{\mu}^{\mathrm{stout}}(n)=\exp (iQ_{\mu}(n))U_{\mu}(n),
\label{eq:stout1}
\end{equation}
where 
$Q_{\mu}(n)$ is Hermitian and traceless and hence in the algebra by design:
\begin{align}
Q_{\mu}(n) &= \frac{i}{2}\left[R_{\mu}^{\dagger}(n)-R_{\mu}(n)-\frac{1}{3}\mathrm{tr}\left(R_{\mu}^{\dagger}(n)-R_{\mu}(n)\right)\mathds{1}\right]\,\ ,\\
R_{\mu}(n)&=C_{\mu}(n)U^{\dagger}_{\mu}(n)\,\ ,\\
C_{\mu}(n)&=\sum_{\nu \neq \mu}\rho_{\mu \nu}(U_{\nu}(n)U_{\mu}(n+\hat{\nu})U^{\dagger}_{\nu}(n+\hat{\mu})+U^{\dagger}_{\nu}(n-\hat{\nu})U_{\mu}(n-\hat{\nu})U_{\nu}(n-\hat{\nu}+\hat{\mu})) \label{eq:stout2b}\,.
\end{align}
Note that within the sum of staples $C_{\mu}(n)$, surrounding the
link $U_{\mu}(n)$, the sum convention is not implied and $\rho_{\mu \nu}$ 
are weights that can be set at will. In our case,
we choose $\rho_{i\nu}=0$ 
and $\rho_{4i}\equiv \rho=1/6$ otherwise.
The value of the
weight was chosen to minimize the one-loop static self-energy after
one smearing iteration. We remark that this is
not necessarily the best possible choice, e.g., in a non-perturbative setting.
We apply only one smearing step to keep the static action local.

\section{Self-energy of a static source}
\label{Sec:deltam}
In this section we introduce our conventions, relate self-energies of static
sources to heavy quark and heavy gluino pole masses, and discuss
the expected renormalon structure.

The triplet and octet self-energies are defined as the lowest
energy eigenvalues of the effective Hamilton operator in temporal
gauge of the sector of Hilbert space of
gauge triplet and octet states with respect
to gauge transformations, applied to a fixed position. 
It is impossible to obtain the continuum limit for these self-energies
(irrespectively of the representation), as these will diverge linearly
with the ultraviolet cut-off.\footnote{In dimensional regularization this object is exactly zero, since the ultraviolet and infrared divergences
(infrared and ultraviolet renormalons)
are regulated by the same factorization scale 
and their sum
vanishes.} Therefore, the value of the static self-energy depends 
on the chosen regulator. Yet,  any hard-cut-off regularization scheme 
is suitable for the following discussion. In this article
we use lattice regularization and write the self-energies in the fundamental 
and adjoint representation in the following way:
\begin{equation}
\label{eq:defmlam}
\delta m=\frac{1}{a}\sum_{n= 0}^{\infty}c^{(3,\rho)}_n\alpha^{n+1}(1/a)\,\mathrm{(fundamental)}, \qquad 
\delta m_{\tilde g}=\frac{1}{a}\sum_{n= 0}^{\infty}c^{(8,\rho)}_n\alpha^{n+1}(1/a)\,\mathrm{(adjoint)}\,,
\end{equation}
where 
$a^{-1}$, the inverse lattice spacing, provides the ultraviolet cut-off. The coefficients $c_n$ depend on the details of the regularization, i.e., on the 
action used and the specific definition of the static propagator.
We only consider the Wilson action~\cite{Wilson:1974sk} but we explore two 
different definitions of the static propagator, with smeared ($\rho=1/6$)
and with the original ($\rho=0$)
temporal links. We label this dependence with a generic $\rho\in\{0,1/6\}$, see
Sec.~\ref{sec:smearing} above. 

One may suspect that the dependence on the regulator might
turn this object uninteresting from the 
theoretical point of view. This is actually not the case
since, for large $n$, $c_n$ becomes regulator independent, universal and 
equal to $r_n/\nu$, the $n+1$ order coefficient of the
perturbative expansion of the pole mass
\be
\label{series}
m_{\OS} = m_{\MS}(\nu) + \sum_{n=0}^\infty r_n \als^{n+1}(\nu)\,\ ,
\ee
up to $\mathcal{O}[\exp(-1/n)]$-terms
(due to subleading renormalons). On an intuitive level this is
clear, as the static energy and pole mass share exactly the
same infrared behavior (up to $\mathcal{O}(1/m)$ corrections),
which should cancel in the difference.

The asymptotic behavior of $r_n$ can be determined assuming that 
the perturbative series is asymptotic and the validity of the OPE.
The running of $\al(\nu)$ is governed by the $\beta$-function
\be
\label{betafunction}
\beta(\al)=\frac{d\al}{d\ln\nu}=-2\alpha\left[
 \beta_0\frac{\al}{4\pi}
+\beta_1\left(\frac{\al}{4\pi}\right)^2
+\beta_2\left(\frac{\al}{4\pi}\right)^3 +\ldots\right]\ ,
\ee
where in our normalization
\be
\beta_0=11\,,\quad\beta_1=102\,,\quad\beta_2^{\MS}=\frac{2857}{2}\,,
\quad\beta_2^{\mathrm{latt}}=-6299.8999(6)\,.
\ee
From $\beta_2$ onwards the coefficients depend on the scheme.
While $\beta_3^{\MS}$ is known~\cite{vanRitbergen:1997va},
in the lattice scheme only $\beta_2^{\mathrm{latt}}$ has been
computed~\cite{Luscher:1995np,Christou:1998ws,Bode:2001uz}. For convenience we define the
constants
\begin{align}
b&=\frac{\beta_1}{2\beta_0^2}\,,\\
s_1&=\frac{\beta_1^2-\beta_0\beta_2}{4b\beta_0^4}
\,,\\
s_2&=
\frac{\beta_1^4 + 4 \beta_0^3 \beta_1 \beta_2 - 2 \beta_0 \beta_1^2 \beta_2 + 
   \beta_0^2 (-2 \beta_1^3 + \beta_2^2) - 2 \beta_0^4 \beta_3}{32 b(b-1)\beta_0^8}
\,,
\end{align}
where only $b$ is scheme-independent.
In a given scheme, the $\Lambda$ parameter is defined as
\begin{align}\label{lambdapa}
\Lambda&=\lim_{\nu\rightarrow\infty}
\nu e^{-\left[
b\ln\left(\frac{\beta_0}{4\pi}\right)
+\int^{\al(\nu)}\!\frac{d\al'}{\beta(\al')}\right]}
\\
\nonumber
&=
\nu e^{-\frac{2\pi}{\beta_0\al(\nu)}}
\left(
\frac{\beta_0\al(\nu)}{4\pi}
\right)^{-b}
\left[
1+s_1 b\frac{\beta_0\al(\nu)}{2\pi}+s_2b(b-1)\left(\frac{\beta_0\al(\nu)}{2\pi}\right)^2+\cdots
\right]
\,,
\end{align}
and we use $\lQ\sim \Lambda_{\MS}$ synonymously with
the size of a typical non-perturbative
binding energy.

A simple scheme- and scale-independent observable is the $B$ meson
mass.
In the heavy quark limit this can be decomposed into the $b$-quark
pole mass $m_{\OS}$ and
the remaining energy from the light quark and gluon dynamics
$\Lambda_B=c_B\lQ$:
\be
m_B=m_{\OS}+\Lambda_B+\mathcal{O}\left(\frac{1}{m_{\OS}}\right)\,.
\ee
This relates to the fundamental representation. For 
the adjoint representation we can think of a heavy gluino attached to gluons:
\be
m_{\tilde G}=m_{{\tilde g},\OS}+\Lambda_H+\mathcal{O}\left(\frac{1}{m_{{\tilde g},\OS}}\right)
\,,
\ee
where $\Lambda_H$ denotes the dynamical contribution of the gluons
(and sea quarks) to the
gluinonium mass.
For both representations the uncertainty of the perturbative
series of the pole mass will 
be of $\mathcal{O}(\lQ)$, the next term in the OPE, since only the sum
of the pole mass and the binding energy has a physical meaning.
This ambiguity results in the successive contributions~$r_n\alpha^{n+1}$ 
to decrease for small orders $n$ down to a minimum at $n_0\sim1/(|a_d|\alpha)$,
where $a_d=\beta_0/(2\pi d)$ with $d=1$. After this order the series starts to diverge, so one neglects the 
higher order contributions and estimates the error by the size of this minimum term,
$r_{n_0}\alpha^{n_0+1}\sim \exp[-1/(a_d\alpha)] \sim \lQ/m$. If the 
perturbative expansion has an ambiguity of order $\lQ^n$ then $d=n$. To quantify this behavior it is
convenient to consider the Borel transform of the above
perturbative series
\be
\label{borel}
m_{\OS} = m_{\MS} + \int\limits_0^\infty\!dt \,e^{-t/\als}
\,B[m_{\OS}](t)
\,,
\quad B[m_{\OS}](t)\equiv \sum_{n=0}^\infty 
r_n \frac{t^n}{n!}\, . 
\ee
The behavior of the expansion
Eq.~\eqref{series} at large
orders is dictated by the closest singularity to the origin of its
Borel transform, which, for the pole mass, is located at
$t=2\pi/\beta_0$, i.e., at $u=1/2$, defining $u=\frac{\beta_0 t}{4 \pi}$.
More precisely,
the behavior of the Borel transform near the
closest singularity at the origin reads
\be
B[m_{\OS}](t(u))=N_m\nu \frac{1}{(1-2u)^{1+b}}
\left(1+s_1(1-2u)+s_2(1-2u)^2+\cdots \right)+(\mathrm{
analytic\; term}),
\ee
where by {\it analytic term} we mean contributions that are expected to be
analytic up to the next renormalon ($u=1$). This dictates the behavior of the perturbative expansion at large orders to be 
\be\label{generalm}
r_n \stackrel{n\rightarrow\infty}{=} N_m\,\nu\,\left(\frac{\beta_0}{2\pi}\right)^n
\,\frac{\Gamma(n+1+b)}{\Gamma(1+b)}
\left(
1+\frac{b}{(n+b)}s_1+\frac{b(b-1)}{(n+b)(n+b-1)}s_2+ \cdots
\right).
\ee
This expression
can be obtained from the procedure employed
in Ref.~\cite{Beneke:1994rs}. The $s_1$-term 
was computed in Ref.~\cite{Beneke:1994rs}, and the $s_2$-term in Refs.~\cite{Pineda:2001zq,Beneke:1998ui}.

As we mentioned, the large-$n$ behavior of $c^{(3,\rho)}_n$
is the same as that of $r_n$ up to $\mathcal{O}(e^{-1/n})$-terms
(due to subleading renormalons). Therefore, using the same scheme
for the expansion parameter $\alpha$, we obtain  
\be\label{generalm2}
c^{(3,\rho)}_n \stackrel{n\rightarrow\infty}{=} N_{m}\,\left(\frac{\beta_0}{2\pi}\right)^n
\,\frac{\Gamma(n+1+b)}{
\Gamma(1+b)}
\left(
1+\frac{b}{(n+b)}s_1+\frac{b(b-1)}{(n+b)(n+b-1)}s_2+ \cdots
\right).
\ee
Note that all the dependence on the regularization details (and in particular on $\rho$) vanishes. The
normalization constant $N_m$ also determines the strength of the renormalon of the singlet static potential, 
through the relation 
\be
\label{eq:nno1}
2N_m+N_{V_s}=0
\,,\ee
since these contributions cancel from
the energy $E(r)=2m+V_s(r)$~\cite{Pineda:1998id,Hoang:1998nz,Beneke:1998rk}.

For adjoint sources we have 
\be\label{cnadj}
c^{(8,\rho)}_n \stackrel{n\rightarrow\infty}{=} N_{m_{\tilde g}}\,
\left(\frac{\beta_0}{2\pi}\right)^n
\,\frac{\Gamma(n+1+b)}{\Gamma(1+b)}
\left(
1+\frac{b}{(n+b)}s_1+\frac{b(b-1)}{(n+b)(n+b-1)}s_2+ \cdots
\right)\,.
\ee
Again, the dependence on the regularization details (e.g.\ on $\rho$)
vanishes,
however, the octet normalization is different: $N_{m_{\tilde g}}\neq N_m$.
Eq.~\eqref{cnadj} corresponds to the renormalon of
gluelump masses (actually $N_{m_{\tilde g}}=-N_{\Lambda}$, where
$N_{\Lambda}$ is the strength of the gluelump renormalon
associated to $\Lambda_H$) and can be related to the pole mass
and adjoint static potential renormalons
through the relation
\be
\label{eq:nno2}
2N_m+N_{V_0}+N_{\Lambda}=0
\,,\ee
since the energy $E(r)=2m+V_o(r)+\Lambda_H$ is renormalon free~\cite{Bali:2003jq}.

To eliminate the unknown normalization constants,
we may consider ratios. In a strict $1/n$ expansion we have\footnote{This equation corrects a mistake in Ref.~\cite{Bauer:2011ws}.}
\begin{align}
\label{cnratioth}
&
\frac{c^{(3,\rho)}_{n}}{c^{(3,\rho)}_{n-1}}\frac{1}{n} =\frac{c^{(8,\rho)}_{n}}{c^{(8,\rho)}_{n-1}}\frac{1}{n} 
\\
&
\quad
=
\frac{ \beta_0}{2\pi}
\left\{1 +\frac{b}{ n} - \frac{b s_1}{n^2}
+\frac{1}{n^3}
\left[b^2s_1^2+b(b-1)(s_1-2s_2)
\right]
+\mathcal{O}\left(\frac{1}{n^4}\right)
\right\}\,\ .
\nn
\end{align}
This expression holds in any representation. It is also
independent of the renormalization scheme used for $\al$.
Keep in mind though that the explicit expression does depend on the
scheme, starting from $\beta_2$. Here we will mainly use $\al_{\mathrm{latt}}$,
where $s_2$ is unknown and
$\beta_2 = \beta_2^{\mathrm{latt}}$, but we will 
also consider the behavior of the perturbative series in the $\MS$ scheme.

Assuming that the coefficients are dominated by the renormalon behavior,
we can determine the order
$n_0+1$ that corresponds to the minimal term
within the pole mass perturbative series. Minimizing $r_n\al^{n+1}$
results in 
\be
\label{minimizing}
(n_0+b)\frac{\beta_0\al}{2\pi}=\exp\left\{-\frac{1}{2(n_0+b)}+\mathcal{O}\left[\frac{1}{(n_0+b)^2}\right]\right\}\,.
\ee
This yields the minimal term
\be
\label{minterm}
r_{n_0}\alpha^{n_0+1}(\nu)=\frac{2^{1-b}\pi}{\Gamma{(1+b)}}\sqrt{\frac{\alpha(\nu)}{\beta_0}}N_m\Lambda
\left[1+\mathcal{O}(\al)\right]\,.
\ee

While it is evident to most readers,
we wish to emphasize that the perturbative
series that defines the pole mass cannot be
resummed (not even in a Borel way).
Therefore, it does not exist in a mathematical sense
and no rigorous numerical value or error can be assigned
to this object.
The most one could do is to define 
a pole mass to a given (finite) order $N+1$,
$m_{\OS}^{(N)}\equiv \sum_{n=0}^{N}r_n\al^{n+1}$, which will then
depend on $N$. By taking $N \sim n_0$ we minimize this
dependence.\footnote{In practice one would round
$N=\mathrm{int}(n_0)$, giving a slightly different value.}
One can then estimate the uncertainty of the sum to
be (see, for instance, the discussion in Ref.~\cite{Sumino:2005cq})
\be
\label{minterm2}
\sqrt{n_0}\,|r_{n_0}|\alpha^{n_0+1}(\nu)=
\frac{2^{3/2-b}\pi^{3/2}}{\Gamma{(1+b)}}\frac{|N_m|\Lambda}{\beta_0}
\,.
\ee
Note that this object is scheme and
scale independent (to the $1/n$-precision that we employed in the
derivation) because, even though the normalizations $N_m$ and $N_{m_{\tilde g}}$
depend on the scheme, the products $N_m\Lambda$
and $N_{m_{\tilde g}}\Lambda$ are scheme-independent.

\section{The Polyakov loop}
\label{Sec:Polya}
We obtain the coefficients $c_n$ from the temporal Polyakov
loop on hyper-cubic lattices. We investigate volumes of $N_T$ lattice points in the time direction
and spatial extents of $N_S$ points.
We choose PBC in time and TBC in all spatial
directions (TBCxyz, see Sec.~\ref{sec:Twist}), eliminating zero modes and improving
the numerical stability. For test purposes, we have performed
additional simulations with PBC in all spatial directions
to $\mathcal{O}(\alpha^{32})$
for a $4^3\times 8$ volume and to lower orders for the specific volumes listed
in the second row of Table~\ref{tab:volumes}.

\begin{table}[h]
\caption{\it 
The $N_S (N_T)$ values of the PBC and TBC runs. The different
geometries are grouped in terms of the orders of the respective
expansions $\mathcal{O}(\alpha^{\nmax+1})$.
\label{tab:volumes}}
\begin{ruledtabular}
\begin{tabular}{ccccc}
&$\mathcal{O}(\alpha^3)$&$\mathcal{O}(\alpha^4)$&$\mathcal{O}(\alpha^{12})$
&$\mathcal{O}(\alpha^{20})$\\\hline
PBC&&4 (4)&&8 (8,10,12,14)\\\hline
\multirow{4}{*}{TBC}&5 (5,6,7,8,10)&4 (5,6,7,8,10,12,16,20,24) &6 (6,8,10,12,16)&7 (7,8)\\
&&&8 (12,16)&8 (8,10), 9 (12)\\
&&&10 (8,12,16,20)&10 (10), 11 (16)\\
&&12 (16,20)&16 (12,16,20)&12 (12), 14(14)
\end{tabular}
\end{ruledtabular}
\end{table}

The Polyakov loop is defined as
\begin{equation}
\label{eq:defpol}
L^{(R)}(N_S,N_T)=\frac{1}{N_S^3}\sum_{\mathbf n}\frac{1}{d_R}
\mathrm{tr}\left[\prod_{n_4=0}^{N_T-1}U^R_4(n)\right]\,,
\end{equation}
where $U^R_{\mu}(n)\approx e^{iA^R_{\mu}[(n+1/2)a]}
\in\mathrm{SU(3)}$ denotes a gauge link in representation
$R$, connecting the
sites $n$ and $n+\hat{\mu}$, $n_i\in\{0,\ldots,N_S-1\}$,
$n_4\in\{0,\ldots,N_T-1\}$.
We implement triplet and octet representations $R$ of
dimensions $d_R=3$ and 8.
The link $U_4(n)$ appears within the
covariant derivative of the static action $\bar{\psi}D_4\psi$,
acting in the following way on a scalar lattice field $f(n)$:
$D_4 f(n)=[U_4(n)f(n+\hat{4})-U_4^{\dagger}(n-\hat{4})f(n-\hat{4})]/(2a)$.
This discretization is not unique and we may substitute
$U_4$ by another gauge covariant connection.
We use singly stout-smeared~\cite{Morningstar:2003gk}
covariant transporters (smearing parameter $\rho=1/6$, see
Sec.~\ref{sec:smearing})
instead of $U_4(n)$
as a second, alternative choice,
to verify the universality of our findings.

We perturbatively expand
the logarithm of the Polyakov loop
\begin{equation}
\label{eq:defP}
P^{(R,\rho)}(N_S,N_T)=-\frac{\ln\langle L^{(R,\rho)}(N_S,N_T)\rangle}{a N_T}
=
\sum_{n=0}^{\infty}c^{(R,\rho)}_n(N_S,N_T)\al^{n+1}
\,,
\end{equation}
in order to obtain the static triplet and octet self-energies
in the infinite volume limit:
\begin{equation}
\label{eq:deltam}
\delta m=\lim_{N_S,N_T\rightarrow\infty}P^{(3,\rho)}(N_S,N_T)\,,
\qquad  
\delta m_{\tilde g}=\lim_{N_S,N_T\rightarrow\infty}P^{(8,\rho)}(N_S,N_T)\ ,
\ee
where (see \Eqre{eq:defmlam})
\be
c^{(R,\rho)}_n=\lim_{N_S,N_T\rightarrow\infty}c_n^{(R,\rho)}(N_S,N_T)
\,.
\end{equation}

The primary objects that we compute are the coefficients
$c_n(N_S,N_T)$. The sets of $c_n(N_S,N_T)$ obtained on different geometries
are statistically independent of one another. However, for a given
volume, different orders $n$ will be correlated.
In computations of
ratios $c_n(N_S,N_T)/c_{n-1}(N_S,N_T)$, as well as in fits, we take
these correlations into account.
NSPT enables us to calculate the coefficients directly, i.e., that 
neither the lattice spacing nor the strong coupling
parameter $\alpha$ enter the simulation.
We have realized a large variety of TBC geometries, listed in
Table~\ref{tab:volumes}, in addition to the PBC test runs.
Each coefficient $c_n$ depends on
$N_S$ and $N_T$ but also on the time step $\epsilon$ of 
the Langevin evolution (see below). In this paper we employ the variant of
the Langevin algorithm introduced
in Ref.~\cite{Torrero:08} and explained in Sec.~\ref{sec:secondorder}, which only quadratically depends on $\epsilon$.
The time series were analyzed following Ref.~\cite{Wolff:2003sm}, 
allowing us to process either single runs or to evaluate
sets of ``farmed out'' Monte Carlo branches. 
Special care was taken to ensure that every individual history
for each order was sufficiently long relative to the
respective autocorrelation time to guarantee a safe error analysis. 
Branches that failed this test were removed from the data analysis.  

Very high statistics
runs were performed up to $\mathcal{O}(\alpha^3)$ and 
$\mathcal{O}(\alpha^4)$,
to check if the coefficients
of logs extracted from the data are in agreement with our
theoretical expectations, to detect signs of ultrasoft
$\ln(N_T/N_S)$-terms (see Secs.~\ref{sec:fse1}
and \ref{sec:FiniteNT} below),
and to compare with results from diagrammatic
perturbation theory.

The bulk of data are obtained up to $\mathcal{O}(\alpha^{12})$ and 
$\mathcal{O}(\alpha^{20})$. We have kept the
Langevin time between two successive measurements fixed,
adjusting $n_{\mathrm{upd}}\approx 56/\epsilon$ where
$n_{\mathrm{upd}}$ is the number of updates
performed in between two measurements.
For the $\alpha^{12}$ and $\alpha^{20}$ runs
between 35000 and 80000 measurements were taken, corresponding
(for $\epsilon=0.050$) to $4\times 10^7$ -- $9\times 10^7$ updates.
The integrated autocorrelation times increase with the
order of the expansion and with the 
lattice volume.
For instance, the integrated autocorrelation time of
$c_0$ varied from 2.4 ($6^4$) to 15 ($16^4$), while
that of $c_{11}$ from $9$ to $30$, in units of $n_{\mathrm{upd}}$.
For our highest order coefficient $c_{19}$ we found
the values $\tau_{\mathrm{int}}\approx 18$ and 29 for
$11^3\times 16$ and $12^4$ lattices respectively.
Since the $\epsilon=0.050$ measurements are separated by
1120 Langevin updates, our largest $\tau_{\mathrm{int}}=30$ value
corresponds to more than 33000 such updates, still leaving
us with a few hundred effectively statistically independent
measurements.

\begin{figure}
\centerline{\includegraphics[width=0.9\textwidth,clip]{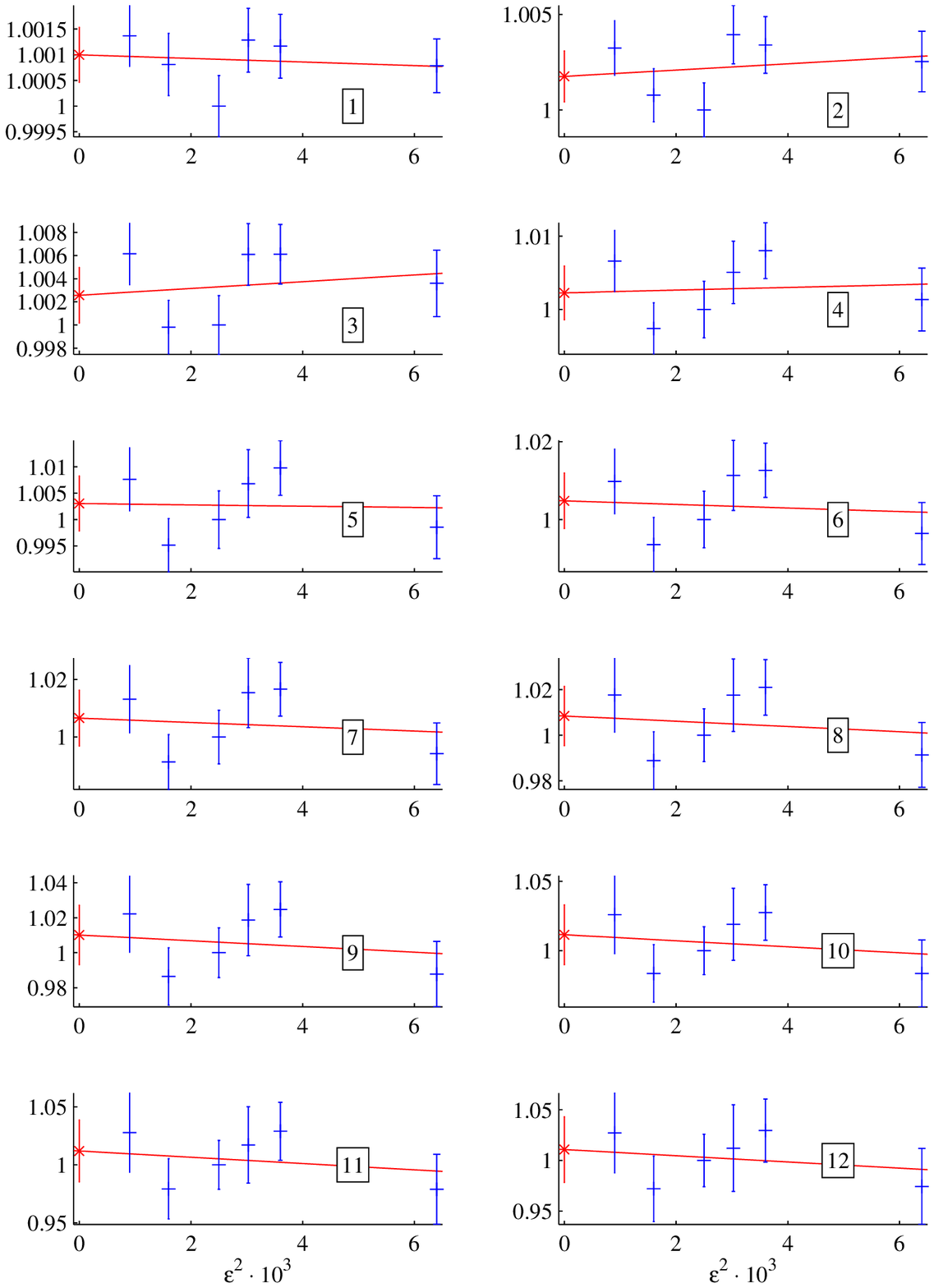}}
\caption{\it  Time step extrapolations of ratios
$c^{(3,0)}_n(16,16;\epsilon)/c^{(3,0)}_n(16,16;0.05)$
(blue symbols) for $\epsilon\in\{0.03,0.04,0.05,0.055,0.06,0.08\}$
at different orders $n+1=1, \ldots, 12$. The left-most red 
symbols are the extrapolated values.\label{fig:L16T16}}
\end{figure}

Quadratic extrapolations in the Langevin time step
to $\epsilon=0$ on a $16^4$ lattice (based on 6 $\epsilon$-values)
and on a $6^4$ lattice (4 $\epsilon$-values) were performed
up to $\mathcal{O}(\alpha^{12})$. 
The $16^4$ extrapolation, normalized to the value obtained at
$\epsilon=0.05$ ($10^3\epsilon^2=2.5$) is shown for
the unsmeared triplet coefficients $c_n^{(3,0)}(16,16)$
in Fig.~\ref{fig:L16T16},
up to $n=11$. A similar picture arises for the $6^4$ volume:
within two standard deviations
the extrapolated values are all found to agree
with the results obtained at $\epsilon=0.05$.
We remark that for both volumes the fit functions are quite flat, 
with very small slopes in $\epsilon^2$.
The same was observed for the smeared and the octet coefficients. 
Based on this experience, the remaining volumes are only
simulated for $\epsilon=0.05$.
However, the time step scaling was only tested within certain
errors that are similar in size as (and in some cases
larger than) the statistical errors obtained for the various geometries.
We obtain a relative
systematic error for each order by adding the statistical error
of the extrapolation
and the difference of the extrapolated value from unity
in quadrature.
For the other geometries this (multiplied by the $\epsilon=0.05$ coefficient)
is then added
in quadrature to the respective statistical error.
For orders larger than
$n+1=12$, we linearly extrapolate in $n$ the systematics
found for orders $n+1\leq 12$,
to obtain an estimate. This procedure is performed not
only for the coefficients themselves but also for ratios
of coefficients. While the systematic time step error is assumed
to be uncorrelated, we keep track of the correlation
between the statistical part of the errors for different orders $n$,
obtained on the same volume.

\section{Finite size effects for $N_T \rightarrow \infty$}

The finite size effects 
of $P(N_S,N_T)$ [Eq.~\eqref{eq:defP}] are well suited
to a theoretical analysis in the limit 
$N_T \rightarrow \infty$ (actually $N_T \geq N_S$ for
most of the geometries that we simulate).
In this limit the self-energy of a static source in
a finite spatial volume is obtained:\footnote{The
discussion in this section applies to any representation
and to smeared and unsmeared 
Polyakov loops.}
\begin{equation}
\delta m(N_S)=\lim_{N_T\rightarrow\infty}P(N_S,N_T)\qquad \mathrm{and} 
\qquad
c_n(N_S)=\lim_{N_T\rightarrow\infty}c_n(N_S,N_T)
\,.
\end{equation}
For large $N_S$, we write 
\be
\label{cnNS}
c_n(N_S)=c_n-\frac{f_n(N_S)}{N_S}+\mathcal{O}{\left(\frac{1}{N_S^2}\right)}\ .
\ee 

\label{sec:fse1}
 \begin{figure}
\centerline{\includegraphics[angle=270,width=.9\textwidth,clip]{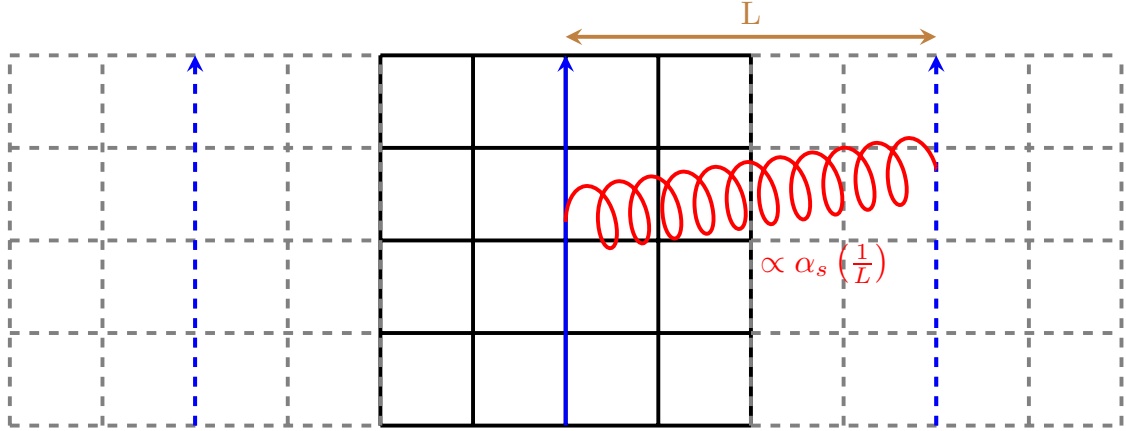}}
\caption{\it \label{fig:alphaNL} Self-interactions with replicas producing
$1/L=1/(aN_S)$ Coulomb terms.}
\end{figure}

The $\mathcal{O}(1/N_S)$ correction originates from interactions with mirror 
images at distances $aN_S$, $\sqrt{2}aN_S$, $\sqrt{3}aN_S$, $2aN_S$, $\ldots$,
see also Ref.~\cite{Trottier:2001vj}. This effectively produces
a static potential between charges 
separated at distances $aN_S$, but without self-energies (the self-energies are included in $\delta m$). As illustrated in Fig.~\ref{fig:alphaNL}, the
scale of such interactions 
is of order $aN_S$ and one may write\footnote{There are some qualifications to this statement that we will detail below.}

\begin{equation}
\label{dmNS}
\delta m(N_S)=\delta m-\frac{1}{aN_S}\sum_{n= 0}^{\infty}f_n\alpha^{n+1}\!\!\left(\left(aN_S\right)^{-1}\right)
+\mathcal{O}{\left(\frac{1}{N_S^2}\right)}
\,.
\end{equation}
Therefore, the coefficient $f_n(N_S)$ is a polynomial of $\ln(N_S)$:
\be
\label{fnNS}
f_n(N_S)=\sum_{i=0}^{n}f_n^{(i)}\ln^i(N_S)\,,
\ee
where $f_n^{(0)} = f_n$ and the coefficients $f_n^{(i)}$ for $i>0$ are entirely determined by  $f_m$ with $m<n$ and $\beta_j$ (see Eq.~\eqref{betafunction}), with $j\leq n-1$. For instance,
\begin{align}
f_1(N_S)&=f_1+f_0\frac{\beta_0}{2\pi}\ln(N_S)\,,
\\
f_2(N_S)&=f_2+\left[2f_1\frac{\beta_0}{2\pi}+f_0\frac{\beta_1}{8\pi^2}\right]
\ln(N_S)+f_0\left(\frac{\beta_0}{2\pi}\right)^{\!\!2}\!\ln^2(N_S)
\,,
\end{align}
and so on.

Starting at $\mathcal{O}(\alpha^4)$, one may expect additional $\mathcal{O}(1/N_S)$ finite size terms. These would arise from infrared singularities of 
certain types of diagrams. The source of these infrared singularities
is similar to the one that results in infrared divergences
of the static potential~\cite{Appelquist:1977es} starting at $\mathcal{O}(\alpha^4)$. In that case these are due to the static
triplet and antitriplet sources, which can arrange themselves
into a singlet or into an octet representation at short distances
in the
pNRQCD~\cite{Brambilla:1999qa} multipole expansion, giving rise to
terms like $\als^4\ln(aN_S\mu_{\mathrm{IR}})/(aN_S)$.  In our case, pairs of
triplet static sources can be arranged into antitriplet and sextet
representations with their mixing mediated through gluons.
At higher orders different representations can arise,
as several mirror images will interact.
Nevertheless, in the limit
$N_T \rightarrow \infty$ in a
finite spatial volume, we do not expect these ultrasoft
logarithms to show up. The reason is that $aN_S$, besides being the
typical momentum transfer between the mirror images, is also the
infrared cut-off of the gluon
momenta so that only logarithmic terms $\sim \ln(N_S/N_S)$ can appear.
Indeed we do not detect any indication of these logs in our numerical data.

Eq.~\eqref{dmNS} can be interpreted in terms of renormalons.
The fact that we can link the $\mathcal{O}(1/N_S)$-term to a
static potential leads to the expectation
that the high order behaviors
of $f_n$ and $c_n$ are dominated by one and the very same renormalon.
This can, e.g., be illustrated 
considering the leading dressed
gluon propagator $D(k)\propto 1/k^2$, where $k_4=0$.
With the (formal) ultraviolet
cut-off $1/a$ and an infrared cut-off $1/(aN_S)$
this can be written as (ignoring lattice corrections),
\begin{align}
\label{cut-offeq}
P&\propto
\int_{1/(aN_S)}^{1/a}\!\!\!\!dk\, k^2 D(k)
\sim\frac{1}{a}\sum_nc_n\alpha^{n+1}\!\left(a^{-1}\right)-\frac{1}{aN_S}
\sum_nc_n\alpha^{n+1}\!\left((aN_S)^{-1}\right)\,,
\end{align}
after perturbatively expanding $D(k)$. 
When re-expressing $\alpha((aN_S)^{-1})$ in terms of
$\alpha(a^{-1})$ we may consider two situations: \\
(a) $N_S>e^n$. In this limit the last 
term of Eq.~\eqref{cut-offeq} is exponentially suppressed in $n$
and the renormalon can directly be obtained
from a large order expansion of $aP$ in powers of $\alpha$.\\
(b) $N_S<e^n$. The last term of Eq.~\eqref{cut-offeq} is
important and the renormalon cancels
order-by-order in $n$. It is easy to visualize the importance of this term in the large-$\beta_0$ limit (see the discussion in 
Ref.~\cite{Pineda:2005kv}).
In this limit one obtains
\begin{align}
\nonumber
\alpha\left(\frac{1}{aN_S}\right)&=\sum_{n\geq 0}
\frac{\ln^n(N_S)}{n!}\left(\frac{d\alpha}{d\ln a}\right)^n\\
&\stackrel{\mathrm{large}\,\beta_0}{\quad=\quad}\sum_{n\geq 0}\left(\frac{\beta_0}{2\pi}\right)^n\ln^n(N_S)\,\alpha^{n+1}\left(a^{-1}\right)\ ,
\end{align}
and therefore the
large-$n$ behaviors of $c_n$ and $f^{(i)}_n(N_S)$ read
\be
\label{cnfni}
c_n \simeq N_m \left(\frac{\beta_0}{2\pi}\right)^nn! \,, \qquad f^{(i)}_n(N_S) \simeq N_m \left(\frac{\beta_0}{2\pi}\right)^n \frac{n!}{i!}
\,.
\ee
This results in the logarithms of Eq.~\eqref{fnNS}
to exponentiate and to cancel the $1/N_S$ suppression.
Therefore, at large $n$, the $f_n/N_S$
terms become numerically as
important as the $c_n$-terms so that
\begin{align}
\label{cnfnNS} 
c_n-\frac{f_n(N_S)}{N_S} &\simeq 
N_m\left(\frac{\beta_0}{2\pi}\right)^nn!\left(1-\frac{e^{\ln N_S}}{N_S}+\frac{1}{N_S}\sum_{i=n+1}^{\infty}\frac{1}{i!}\ln^iN_S\right)
\nn
\\
&=
\frac{1}{N_S}N_m \left(\frac{\beta_0}{2\pi}\right)^n n!
\sum_{i=n+1}^{\infty}\frac{1}{i!}\ln^i(N_S)\,.
\end{align}
Note that the renormalon ($n!$ behavior) actually cancels
in the difference and that the infinite sum above is a convergent series
($N_S$ minus a finite sum).

In present-day numerical
simulations, including ours, $N_S<e^n$, and the term $f_n(N_S)/N_S$ needs to be
taken into account, in combination with $c_n$. A similar
phenomenon was numerically observed for
the static singlet energy $E(r)=2m+V(r)$~\cite{Pineda:2002se,Bali:2003jq}.
This teaches us that to correctly identify
the renormalon structure of $\delta m$, it is compulsory to 
incorporate the $1/N_S$ corrections.
So far, in studies of high order perturbative expansions of the plaquette
the corresponding finite size terms have been neglected.
As we will see, our fits indeed yield $f_n\simeq c_n$ for large $n$,
in clear support of the renormalon dominance picture.

In the $N_T \rightarrow \infty$ limit, and up to $\mathcal{O}(1/N_S^2)$ effects, the fit function for $c_n(N_S)$ depends on $c_n$, $f_n$ and the $\beta$-function
coefficients $\beta_i$ with $i<n$. 
In the lattice scheme only $\beta_0$, $\beta_1$ and $\beta_2$ are
known. The effects of higher $\beta_j$ start 
at $\mathcal{O}(\alpha^5)$. One may try to fit these,
together with the $c_n$ and $f_n$ but their contribution cannot
be resolved by the present precision of the data. This produces
some uncertainty in our parametrization that we will add to the error.
The reason we can neglect higher $\beta_j$ in a controlled way is
because the associated uncertainty quickly becomes negligible at high
orders, once the behavior of the coefficients $f_n$ starts
to be governed by the $d=1$ renormalon. This statement can be quantified, since 
we know the large-$n$ behavior of the $f^{(j)}_n$.
Let us first consider the large-$\beta_0$ limit.
Assuming renormalon dominance for the coefficients
$f_n$, we would have (see Eqs.~\eqref{cnfni} and \eqref{cnfnNS})
\be
\label{70}
f_n(N_S)=N_m \left(\frac{\beta_0}{2\pi}\right)^nn!
\sum_{i=0}^{n}\frac{1}{i!}\ln^i(N_S)
\,.
\ee
Note that terms containing higher powers of $\ln(N_S)$ (with $i \lesssim n$)
are suppressed by factors $\sim 1/n!$ and can be neglected at large $n$. 
Therefore, for large $n$, the $n!$ factorial overcomes the large logs. 
This suppression also holds beyond the large-$\beta_0$
limit.\footnote{In any case, 
for any given $n>3$, the coefficients $f^{(n),(n-1),(n-2)}_n$ are
completely determined by the renormalization group analysis and 
the coefficients $f_1$, $f_2$ and $f_3$.}
However, we may worry about terms with small powers of
$\ln(N_S)$ ($i \sim 1$ in Eq.~\eqref{70}). In this case there is
no factorial suppression, but the inclusion of $\beta_1$ and $\beta_2$
can still be done in a controlled way. Including the running
associated to $\beta_1$ produces $1/n$ suppressed corrections
to Eq.~\eqref{cnfni}
while $\beta_2$ results in $1/n^2$ suppression and so on.
For instance in the case of $f_n^{(1)}$ we have 
\be
f_n^{(1)}=N_m\,\nu\,\left(\frac{\beta_0}{2\pi}\right)^n
\,\frac{\Gamma(n+1+b)}{\Gamma(1+b)}
\left[
1+\frac{b}{(n+b)}\frac{n-1}{(n-1+b)}+\mathcal{O}\left(1/n^2\right)
\right]\,.
\ee
To check this assumption and to justify
the truncation at $\beta_2$
we have performed separate fits including $\beta_j$
for $j\leq 0, 1$ and 2 (see Sec.~\ref{Sec:fits}).

Finally note that the validity of the
above discussion is unaffected by any
renormalon (or other singularity)
related to the $\beta$-function coefficients, as this would
correspond to a higher
dimension, i.e., $u>1/2$ (if it existed at all).

\section{$1/N_T$- and subleading $1/N_S$-corrections}
\label{sec:FiniteNT}
While most of our geometries satisfy $N_T \geq N_S$, the $N_T$
dependence may still be sizable and cannot completely be
neglected \emph{a priori}. This may necessitate a combined expansion in
powers of $1/N_S$ and $1/N_T$. The leading order (LO) correction in
$1/N_S$ has been discussed in the previous section. Incorporating
finite $1/N_T$ effects does not affect the renormalon structure
nor the main conclusions of that section.
The only subtlety that we need to revisit
are the ultrasoft effects.
In principle, a dependence on $N_T$ may appear, starting
at $\mathcal{O}(\alpha^4)$. Nevertheless, only in the limit $N_T \ll N_S$ 
do we expect large logs of the type $\ln(N_S/N_T)$, as in this
limit $1/(aN_T)$ may act as the infrared regulator. Still our geometries
are far off this limit. One may consider an interpolating phenomenological
function
like $\ln[N_S/(N_T+N_S)]$ between the $N_T \ll N_S$ and $N_S \ll N_T$
limits, yet the data do not seem to require these terms. 
We also stress that these terms are subleading from the renormalon
point of view ($d=3$). Therefore, in our final fit function we
will not introduce them.

 \begin{figure}
\centerline{\includegraphics[width=0.9\textwidth]{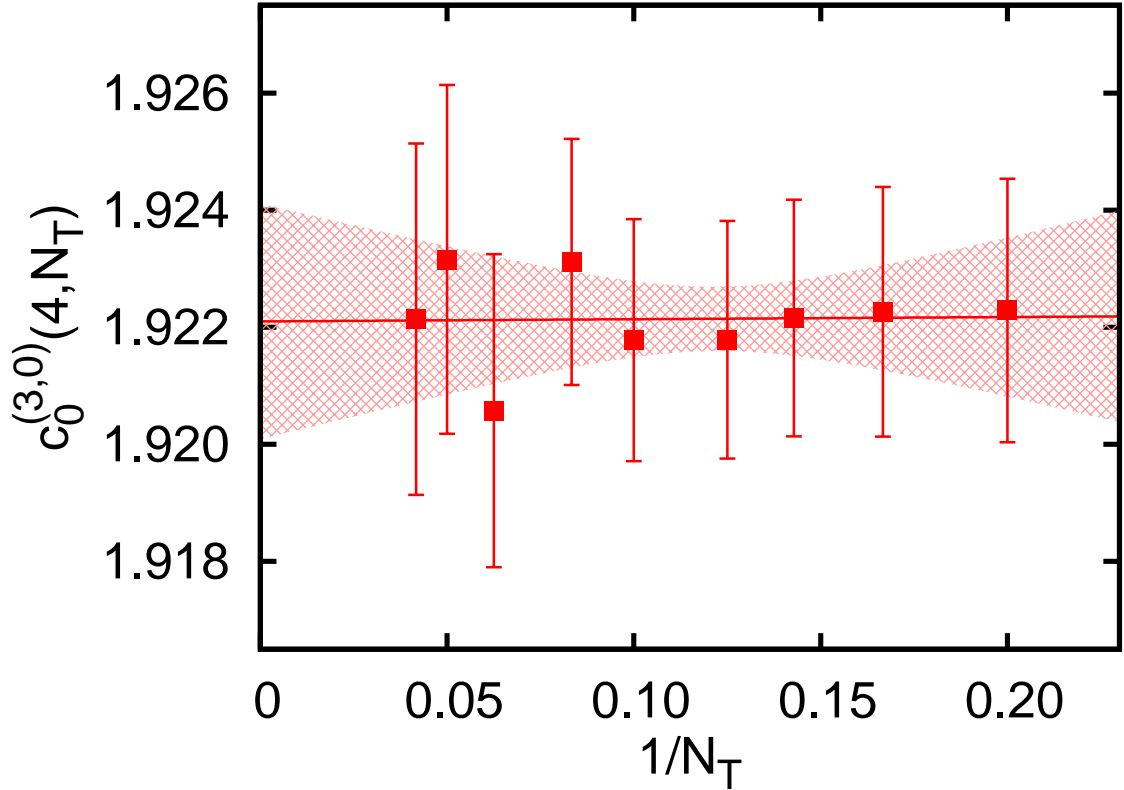}}
\caption{\it \label{fig:Alpha_L4_T} $c^{(3,0)}_0(4,N_T)$ as a
function of $1/N_T$ compared to a linear fit. The linear term is
clearly zero within errors. The fit gives
$c^{(3,0)}_0(4)=1.9221(20)$,
to be compared to $1.92253$ from DLPT, Eqs.~\protect\eqref{c0pert}--\protect\eqref{c00pert}.}
\end{figure}

We now study possible power-suppressed $1/N_T$ effects. First we
consider the low orders in perturbation theory. At
$\mathcal{O}(\alpha)$ the fit function with finite (but large)
$N_T$ can be obtained with DLPT.
No dependence on $N_T$ is found. This is also confirmed by our
explicit computation of $c_0$ with NSPT. We illustrate this
for $N_S=4$ in Fig.~\ref{fig:Alpha_L4_T}. A similar picture applies
to the other values of $N_S$.
The leading terms in $1/N_S$ can also be determined using DLPT. Writing
\be
\label{c0pert}
c^{(3,0)}_0(N_S,N_T)=c^{(3,0)}_0(N_S)=c^{(3,0)}_0-\frac{f^{(3,0)}_0}{N_S}-\frac{v^{(3,0)}_0}{N^3_S}+\mathcal{O}\left(\frac{1}{N_S^4}\right)
\,,\ee
we obtain for unsmeared coefficients and TBCxyz boundary conditions
\begin{align}
\label{c30DLPT}
c^{(3,0)}_{0,\mathrm{DLPT}}&=2.1172743570834807985970\ldots\,,\\
\label{f30DLPT}
f^{(3,0)}_{0,\mathrm{DLPT}}&=0.76962563284(2)\,,\\
v^{(3,0)}_{0,\mathrm{DLPT}}&=0.14932(3)\,.\label{c00pert}
\end{align}
Note that DLPT predicts the absence of an $\mathcal{O}(1/N_S^2)$
term at this order. 
The above result also applies to the adjoint source, substituting
$c^{(8,\rho)}_{0}(N_S,N_T)=C_A/C_F\, c^{(3,\rho)}_{0}(N_S,N_T)$, where
$C_A/C_F=2N_c^2/(N_c^2-1)=9/4$.
We remark that $f_0$ and $v_0$ depend
on the boundary conditions, whereas $c_0$ does not.

As we already mentioned, the finite volume
$c^{(3,0)}_0(N_S)=c^{(3,0)}_0(N_S,N_T)$ depend on the boundary
conditions. It has previously been computed with PBC, originally
in Ref.~\cite{Heller:1984hx}, where intermediate semi-analytic
expressions can be found, and in Refs.~\cite{Nobes:2001tf,Trottier:2001vj}
where also TBCxyz and TBCxy boundary conditions were analyzed.
No time dependence was found in either case. This absence of
a time dependence at $\mathcal{O}(\alpha)$ fits
with the spectral picture. 
The infinite volume coefficient was most precisely
computed in Ref.~\cite{Necco:2001xg}. 
Our determination of $c_0$ agrees with the previous results.

At $\mathcal{O}(\alpha^2)$ we start to encounter a dependence on $N_T$.
DLPT also gives us information on the coefficient 
$c_1(N_S,N_T)$. In this case we have only computed the infinite volume
limit for the unsmeared coefficient using the code
of Ref.~\cite{Bali:2002wf} in DLPT:
\be
\label{c130}
c^{(3,0)}_{1,\mathrm{DLPT}}=C_F/C_A\, c^{(8,0)}_{1,\mathrm{DLPT}}=11.1425(25)\,.
\ee
$c^{(3,0)}_1$ has been computed previously in a less controlled way
using finite Wilson
loops~\cite{Martinelli:1998vt}, resulting in the value $c^{(3,0)}_1=11.152$.
In Ref.~\cite{Nobes:2001tf} agreement with this value was reported.
Beyond $\mathcal{O}(\alpha^2)$ there exist no DLPT results.

\begin{figure}
\centerline{\includegraphics[width=0.48\textwidth]{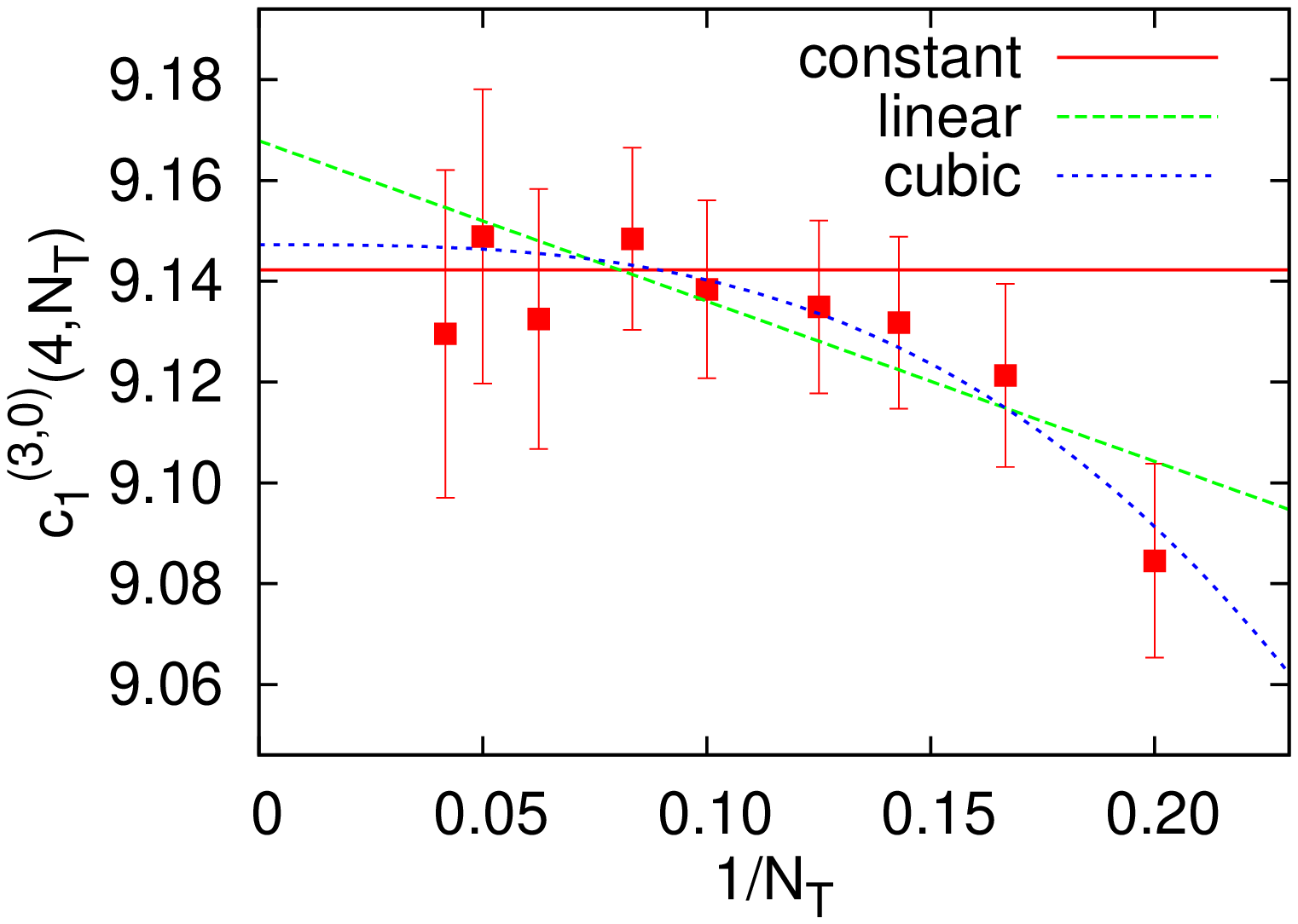}~
\includegraphics[width=0.48\textwidth]{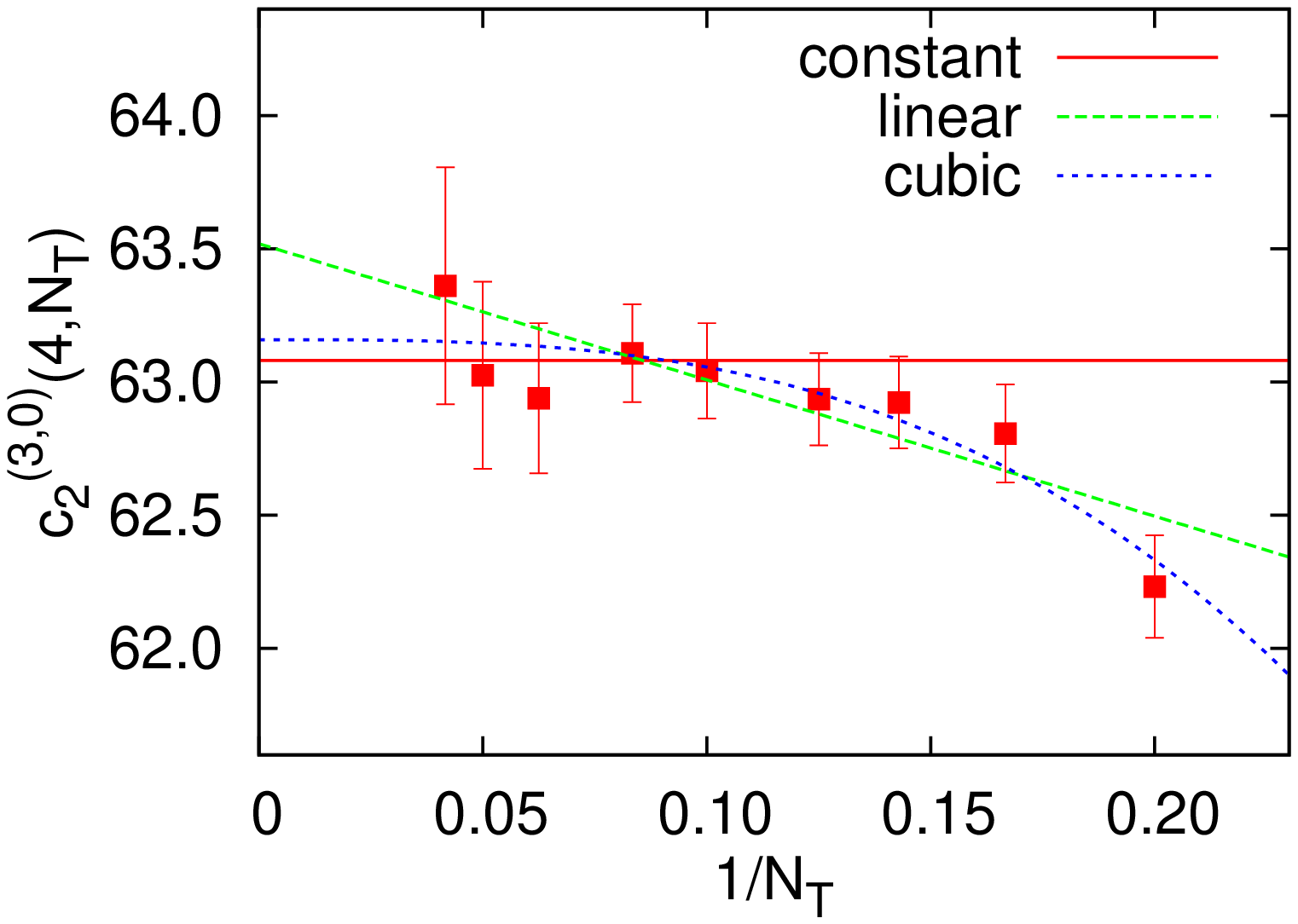}}
\centerline{\includegraphics[width=0.48\textwidth]{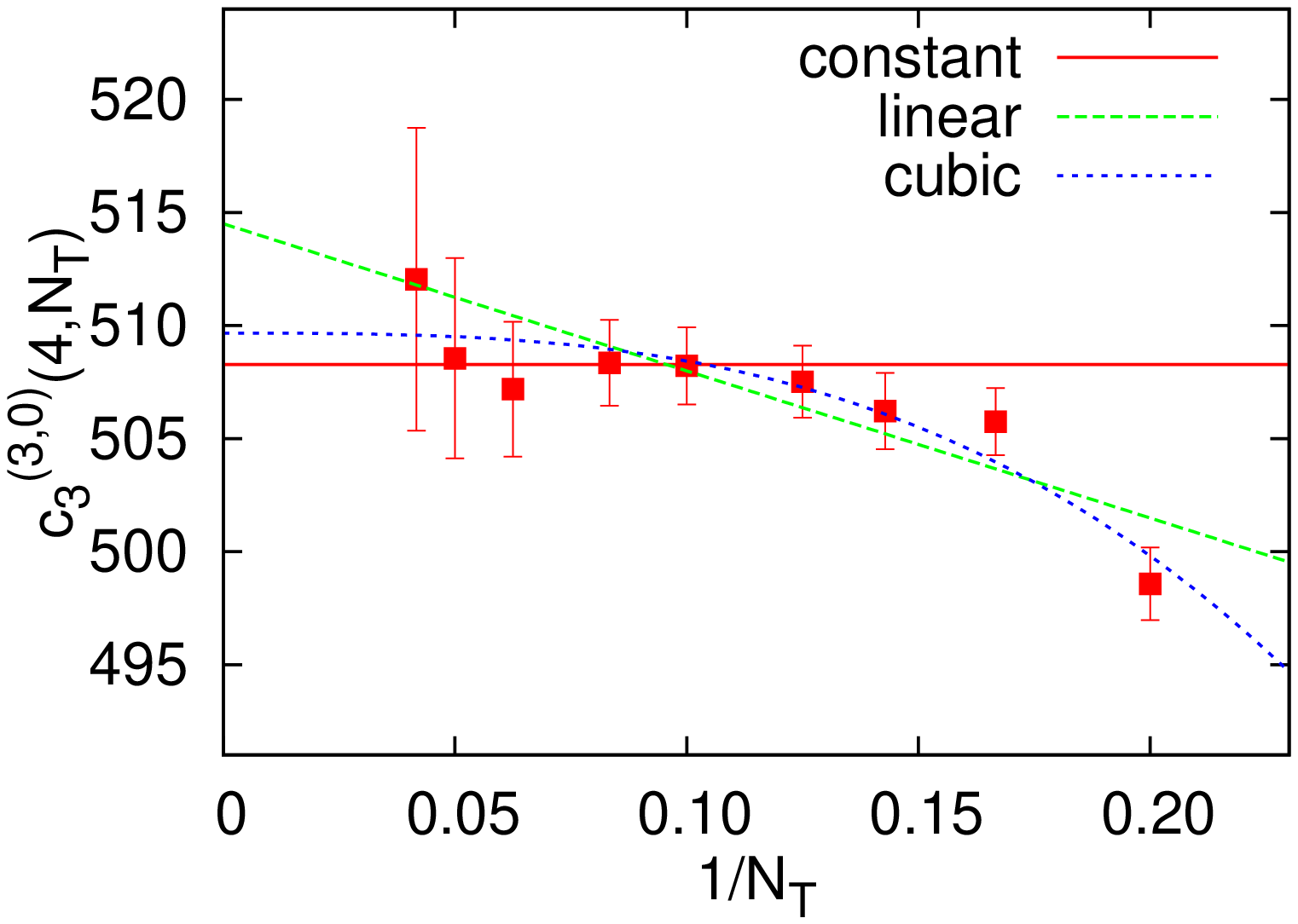}}
\caption{\it \label{fig:Alpha2_L4_T} 
$c^{(3,0)}_{1,2,3}(4,N_T)$ as a function of $1/N_T$, in comparison to a constant plus linear fit, 
a constant plus cubic fit, and a constant fitted only to the $N_T>10$ points.}
\end{figure}

Next we address NSPT data for $n\geq 1$. We wish to understand
the $N_T$ dependence for 
$N_T > N_S$. For this analysis the simulations
to $\mathcal{O}(\alpha^4)$ at $N_S=4$ up to the very high $N_T=24$ turn out
to be particularly useful. The results are
shown in Fig.~\ref{fig:Alpha2_L4_T}. Note that in these cases
the error bars are
dominated by the finite
Langevin timestep systematics.
The $n=1,2,3$ results all  show the same qualitative behavior.
For $N_T\geq 10$
the data are constant within errors, and linear fits result in slopes that
are compatible with zero within two standard deviations.
For $N_T$ smaller than 10 we start to see a bending in $1/N_T$,  
which we parameterize by a $1/N_T^d$ function. Large powers of $d$ 
are favored by the fit. The specific power is difficult to determine. 
We find a $1/N_T^5$ fit to best describe the data,
though only marginally better than a $1/N_T^3$ fit.
Linear $1/N_T$ fits, however, 
are unsatisfactory as we can see in Fig.~\ref{fig:Alpha2_L4_T}. We can
compare the $c^{(3,0)}_{1,2,3}(4,\infty)$-values obtained
averaging $N_T>10$ data vs.\ performing $1/N_T^3$ fits:
$9.142\pm 0.012$ vs.\ $9.147\pm 0.010$, 
$63.08\pm 0.13$ vs.\ $63.16\pm 0.10$ and
$508.3\pm 1.5$ vs.\ $509.7\pm 1.0$.
Indeed, within our present accuracy,
the large-$N_T$ data are in agreement with the extrapolation.
The same also holds for the smeared and octet data sets.

\begin{figure}
\centerline{\includegraphics[width=0.8\textwidth]{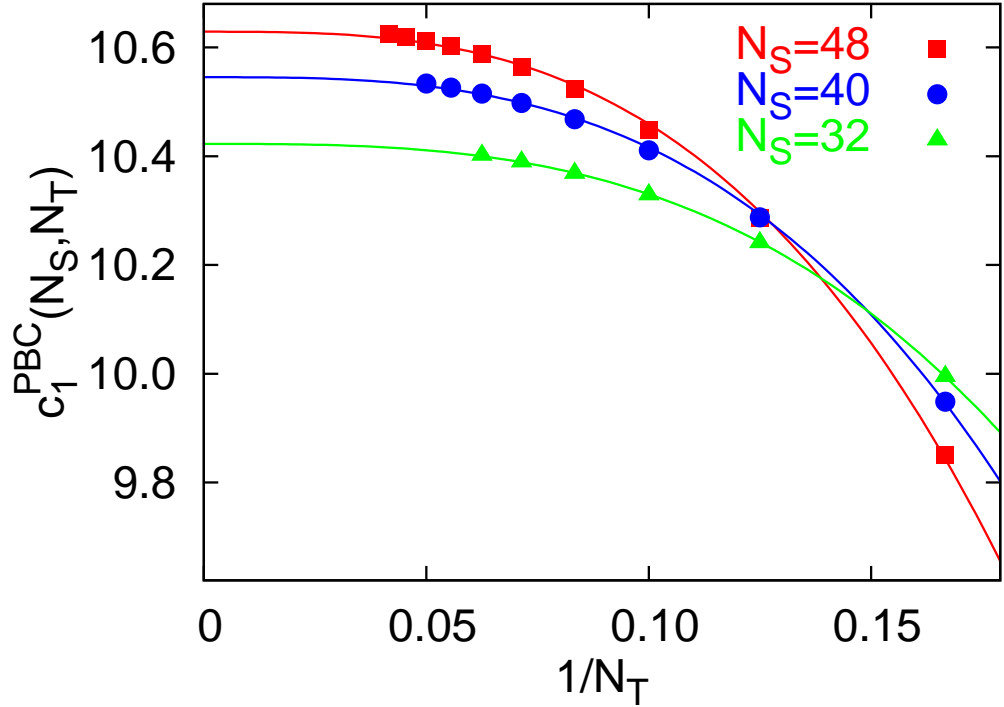}}
\caption{\it \label{fig:PBC}
$c^{(3,0)}_1(N_S,N_T)$ from DLPT obtained on volumes with PBC.
The fitted curves are constant plus cubic ($1/N_T^3$).}
\end{figure}

We found
$1/N_T^5$ and $1/N_T^3$ fits to also work well for other $N_S$-values
(though in these cases we have less data points and therefore less
conclusive results). Irrespectively of the power $d$,
we observe the coefficient of the $1/N_T^d$-term
to increase roughly linearly with $N_S$. From
this phenomenological analysis we conclude that the $1/N_T$ effects
effectively count as $N_S/N_T^3 \sim 1/N_T^2$. Moreover,
we find the coefficients of these terms to be  numerically small.
In order to confirm this phenomenological counting, 
we have also explored the region $N_T \leq N_S/2$ for $n=1$ using the
PBC DLPT
formulae of Ref.~\cite{Heller:1984hx} that only
apply to $N_T<N_S$.
In this case for very large volumes ($N_S\geq 32$), 
we indeed found a $(a+b N_S)/N_T^3$ parametrization
to work well, see Fig.~\ref{fig:PBC}.
This means that the $1/N_T$ effects are clearly subleading, compared to
the $1/N_S$ effects that we incorporate in our fit and
also subleading relative to the unknown $1/N_S$
effects starting at $\mathcal{O}(\alpha^5)$,
due to $\beta_3$. 

While these analyses strongly indicate that the
$1/N_T$ effects decay rapidly with $N_T$, the specific functional form
is not exactly known. Therefore, our analysis strategy will be to take
$N_T \geq\max(N_S,11)$ so that the $1/N_T$ effects can safely be neglected.
In this way we loose data and statistics but avoid any
bias from assuming a particular functional
form. To estimate the cut-off systematics we
then vary $N_T$ and also
consider different trial fit functions. We discuss this issue
further in the next section.

Subleading effects of $\mathcal{O}(1/N_S^2)$ would
be obscured by the unknown (logarithmically modulated)
$1/N_S$ effects from higher $\beta$-function coefficients.
Therefore we will not consider these.

We conclude with a discussion of lattice artifacts.
Formally, we may introduce an anisotropy $a_t\neq a_s$. In this case
the lattice action, that is invariant under time or parity reversal,
agrees with the continuum action up to $\mathcal{O}(a_t^2,a_s^2)$-terms.
The temporal and spatial lattice extents in physical units are given by
$a_tN_T$ and $a_sN_S$, respectively, so that the
only dimensionless combinations consistent with the leading
order lattice artifacts are $a_t^2/(a_tN_T)^2=1/N_T^2$ and $1/N_S^2$.
Therefore, within perturbation theory, where we cannot dynamically generate
additional scales, the LO lattice artifacts are
indistinguishable from $\mathcal{O}(1/N_T^2,1/N_S^2)$ finite
size effects which, as discussed above, are beyond our present level of
precision. 

\section{Simulation Results}
\label{Sec:fits}
In this section we obtain the infinite volume coefficients of
the expansions of four different self-energies: for
fundamental and adjoint sources and using static actions
with smeared and
unsmeared time derivatives. We compare their large order
behavior with theoretical expectations, and determine the
leading renormalon normalizations $N_m$ and $N_{m_{\tilde g}}$. We then
convert the results
into the $\overline{\mathrm{MS}}$ scheme, using different methods,
and estimate $\beta_3^{\mathrm{latt}}$.

\subsection{Infinite volume coefficients}
\label{Sec:fitscoeff}
Below we determine the infinite volume coefficients $c_n^{(R,\rho)}$
defined in Eq.~\eqref{eq:defmlam}. 
Our default fit function for $c^{(R,\rho)}_n(N_S,N_T)$ (see Eq.~\eqref{eq:defP}) is defined
in Eqs.~\eqref{cnNS}--\eqref{fnNS}, and depends on the fit parameters $c^{(R,\rho)}_n$ and
$f^{(R,\rho)}_j$ with $j\leq n$. This last dependence introduces a correlation between different $n$-valued coefficients, 
which we take into account by simultaneously fitting\footnote{All the
global fits to the $c_n(N_S,N_T)$ data have been double checked by two
different program implementations using both Maple and Mathematica.} 
all $c_n(N_S,N_T)$ to data up to a given order $\mathcal{O}(\al^{\nmax+1})$. 
By default $\nmax+1=20$.
As a sanity check,
we have also fitted each order $n$ independently with
two fit parameters $c_n$ and $f_n$, keeping the
$f_j$-values that were obtained at previous orders $j<n$ fixed.
Since this iterative method does not take account of all correlations, the resulting statistical errors
and $\chi^2$-values are not reliable.
Nevertheless, these fits yield similar central values, illustrating
that the low order coefficients are only mildly affected
by higher order data.
In the following we will only use the results of the global fits.

To ensure that $1/N_T$ effects can be neglected
we restrict our fits to
$N_T \geq \max(N_S,\nu_T)$ with $\nu_T=11$.
In addition we restrict $N_S \geq \nu_S$, varying $\nu_S$ to
explore the validity range of Eq.~\eqref{cnNS}. Our ``thermometer''
for this will be to obtain acceptable $\chi^2/N_{\mathrm{DF}}$-values and good
agreement with $c_1^{(3,0)}$ and $c_1^{(8,0)}$ from DLPT, Eq.~\eqref{c130}.
We find that including small volumes improves the quality of the
fits: the values of $c_1^{(R,0)}$ tend towards the expectations, and $\chi^2/N_{\mathrm{DF}}$, as well as the errors, are reduced. We illustrate this behavior in Table~\ref{tab:c0}. We have observed the same behavior for 
different values of $\nu_T$ around 11, and also for the octet and/or smeared perturbative series. 
Therefore, our default setting will be $\nu_S=4$. 

\begin{table}[h]
\caption{\it 
$\chi^2/N_{\mathrm{DF}}$, $c_1^{(3,0)}$ and $c_{19}^{(3,0)}$ for different
values of $\nu_S\leq N_S$ ($N_T\geq\max(N_S,11)$). The $n=0$ values
were fixed to the DLPT result. Otherwise the $\chi^2/N_{\mathrm{DF}}$-values
come out even smaller: 1.570, 1.322, 1.209 and 1.152 respectively, whereas
the coefficients barely change. The DLPT expectation is
$c_1^{(3,0)}=11.1425(25)$.\label{tab:c0}}
\begin{ruledtabular}
\begin{tabular}{c|cccc}
$\nu_S$&9&7&6&4\\\hline 
$\chi^2/{N_{\mathrm{DF}}}$&1.701& 1.431& 1.309 & 1.263\\
$c_1^{(3,0)}$& 11.120(33)& 11.124(25)& 11.122(17) & 11.136(11)
\\
$c_{19}^{(3,0)}/10^{23}$&3.919(73)& 3.995(55)& 
4.108(36)& 4.118(36)
\end{tabular}
\end{ruledtabular}
\end{table}

\begin{figure}[htb]
\centerline{\includegraphics[width=0.9\textwidth,clip]{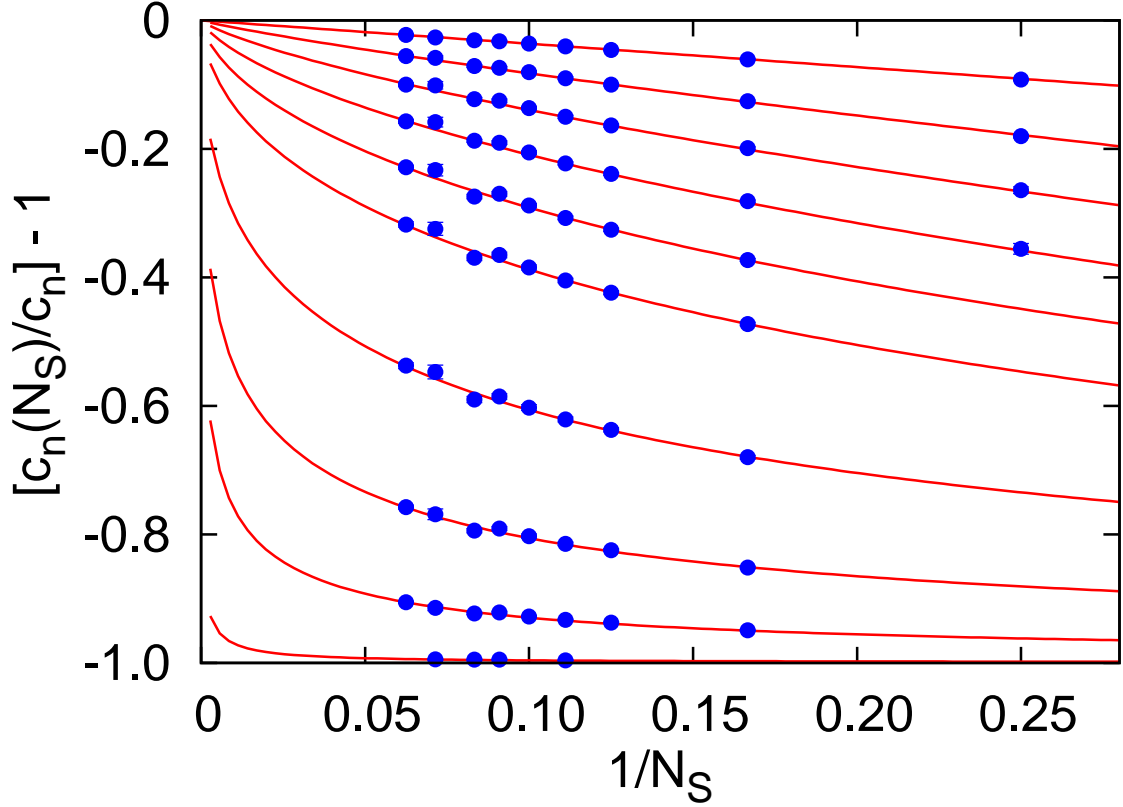}}
\caption{\it
$c^{(3,0)}_n(N_S)/c^{(3,0)}_n-1$  for $n\in\{0,1,2,3,4,5,7,9,11,15\}$ (top to bottom). For each value of $N_S$ 
we have plotted the data point with the maximum value of $N_T$.
The error bars are invisible on the scale of the figure.
The curves represent the global fit. For $n=0$ the DLPT
prediction $-(1/N_S)f^{(3,0)}_{0,\mathrm{DLPT}}/c^{(3,0)}_{0,\mathrm{DLPT}}$
of Eqs.~\protect\eqref{c30DLPT} and \protect\eqref{f30DLPT} is shown
(straight line).
\label{fig:NLrunning}}
\end{figure}

\begin{figure}[htb]
\centerline{\includegraphics[width=0.9\textwidth,clip]{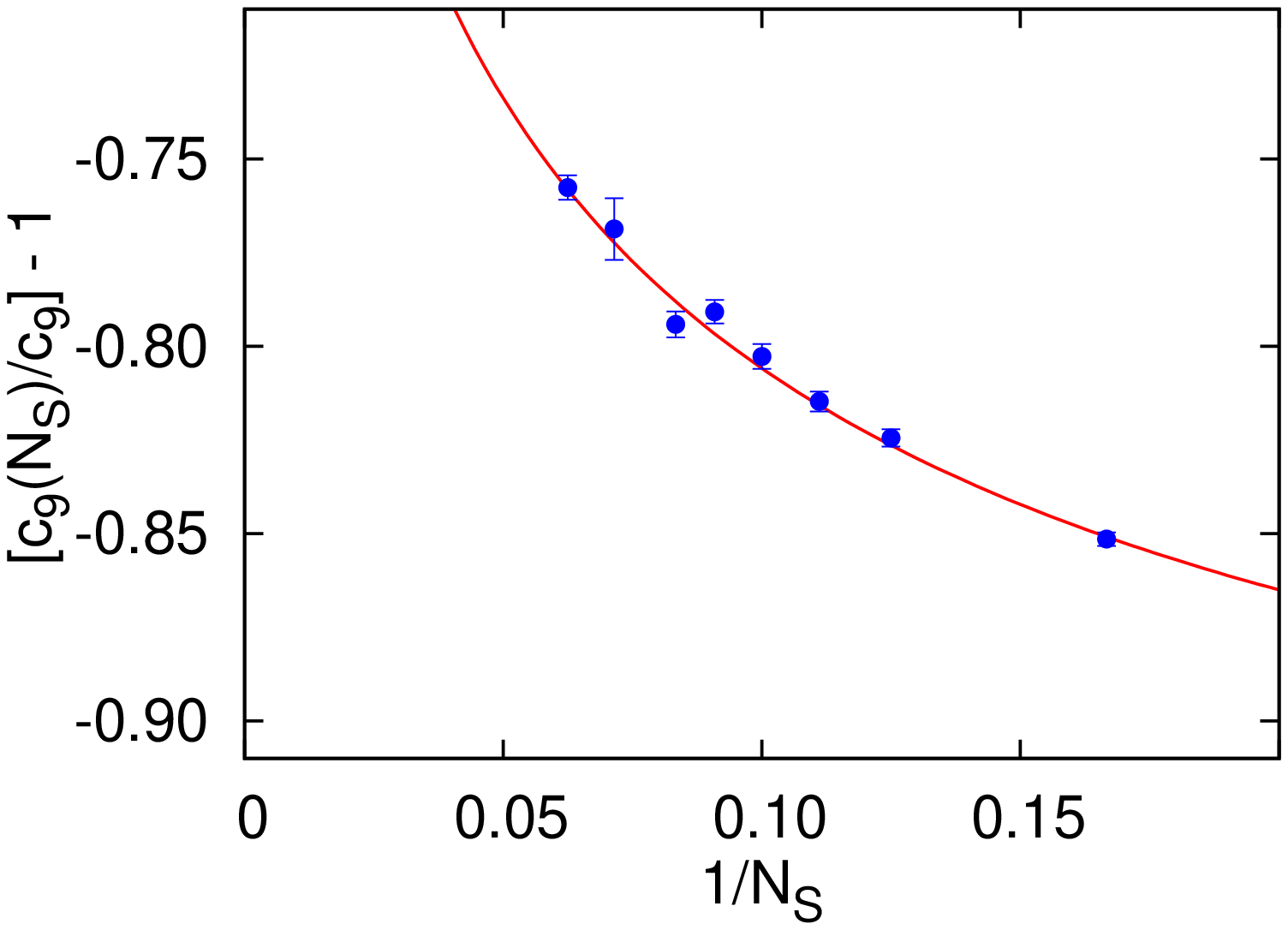}}
\caption{\it
Zoom of Fig.~\protect\ref{fig:NLrunning} for $n=9$.
\label{fig:NLrunning9}}
\end{figure}

The leading parametrical uncertainty stems from the unknown
$1/N_S$ effects associated to higher order terms in the
$\beta$-function: $\beta_3$, $\beta_4$ etc., which will
start affecting the fit at orders $n+1\geq 5$. As long
as all singularities of the lattice $\beta$-function in the Borel plane
are further away than $u=d/2=1/2$ from the origin
(which is the case), these higher $\beta_i$ coefficients
will not affect the leading renormalon
behavior. Nevertheless, there can be an impact at
intermediate orders. To study this, for each
$c_n^{(R,\rho)}(N_S,N_T)$ we perform
three different fits, setting $\beta_1=\beta_2=0$ 
($\beta_0$), setting only $\beta_2=0$ ($\beta_{0,1}$), and
using all the known coefficients ($\beta_{0,1,2}$).
The resulting $c_n^{(3,0)}$ are displayed in Table~\ref{tab:cnCoeffs}
of the Appendix.
The results between the $\beta_0$ and $\beta_{0,1}$ fits start to
deviate from each other significantly at $n=4$ while
$\beta_{0,1,2}$ becomes statistically
distinguishable from $\beta_{0,1}$ starting
around $n=9$. At $n=19$ there is a 25~\% variance between
the $\beta_0$ and $\beta_{0,1,2}$-fits. The convergence pattern
is sign alternating. The picture is similar
for the smeared and the octet results.
We will take the difference between the $\beta_{0,1}$ and
$\beta_{0,1,2}$ results as an estimate of the error from subleading
terms in the $\beta$-expansion. This is our dominant
source of systematic error, by far exceeding, e.g., our statistical errors.
We remark that switching off the running altogether ($\beta_i=0$)
yields a bad $\chi^2/N_{\mathrm{DF}}=3.167$ (with $n=0$ fixed from DLPT) and
a value of $c^{(3,0)}_1$ that is by about 20 standard
deviations away from the DLPT result. Once the running is introduced 
into the parametrization of the finite size effects,
these quickly and unavoidably (see Sec.~\ref{sec:fse1})
grow in size, resulting in large cancellations
with the coefficients $c_n$. We illustrate the importance of
this effect in Fig.~\ref{fig:NLrunning}, where we compare
the fitted parametrization to the unsmeared triplet data
on $c_n(N_S)/c_n-1$ for various $n$
(this also illustrates the quality of the fit).
Note that the curvatures, i.e.\ the deviations from straight lines,
are due to the renormalization group running of the $1/N_S$ coefficients.
The data clearly show the expected curvature. To illustrate this
better, we enlarge the $n=9$ curve in Fig.~\ref{fig:NLrunning9}.

Next, we estimate the error associated to the $N_T$-range dependence that
we have not accounted for in our fits. Our data run 
over a large variety of lattice volumes with different $N_T$-values.
Our cut-off $N_T \geq 11$ eliminates a significant fraction of
lattice geometries. However, we can still benefit from these
discarded volumes, as they allow us to estimate the systematics
associated with our choice of cut-off. We follow two strategies:
i) we vary the cut-off $\nu_T$. We display $\nu_T=9$ results in
Table~\ref{tab:NTError}.
Reducing the cut-off increases the $\chi^2/N_{\mathrm{DF}}$-values,
since the $1/N_T$ curvature is not built into our parametrization.
 Other than this
there is good agreement with our $\nu_T=11$ $\beta_{0,1,2}$ fits
of Table~\ref{tab:cnCoeffs}.
ii) We introduce a 
$N_T$-dependent term into the fit function in the following way\footnote{In
Ref.~\cite{Bauer:2011ws} we employed a 
different parameterization of the $N_T$-dependence.
The fit yielded similar results to those found here but
using two extra parameters per order.}
\be
\label{cnNSNT}
c_n(N_S,N_T)=c_n-\frac{f_n(N_S)}{N_S}+\frac{v_n(N_S)}{N_T^d}
\,,
\ee 
and fit to all our volumes ($\nu_T=5$).
We have explored different values of $d$ and different
parametrizations of $v_n(N_S)$. 
In Sec.~\ref{sec:FiniteNT} the low $n$ $v_n(N_S)$ coefficients were found
to increase with $N_S$. Global fits also favor this behavior. 
Therefore, we consider two fit functions:
ii.a) $v_n(N_S)/N_T^d$ where we construct
$v_n(N_S)$  in analogy to the $f_n(N_S)$-term, using the renormalization group running
of previous orders with just one new fit parameter $v_n=v_n^{(0)}$
at each order.
ii.b) $\tilde{v}_nN_S/N_T^d$, assuming a linear dependence
of this term on $N_S$. We now vary $d$. We take $d=2$ for the
ii.a) fit, as we obtain a good $\chi^2/N_{\mathrm{DF}}$-value 
and agreement with $c_{1,\mathrm{DLPT}}$ within one standard
deviation. Varying $d$ increases $\chi^2/N_{\mathrm{DF}}$ and
deteriorates this agreement. 
We take $d=3$ for the ii.b) fit, as it yields a good
$\chi^2/N_{\mathrm{DF}}$-value and also perfect agreement
with $c_{1,\mathrm{DLPT}}$. $d=2$ results in a difference
between the fitted value of $c_1$ and $c_{1,\mathrm{DLPT}}$ of
several standard deviations, while $d=4$ and $d=5$ reduce
the quality of the global fit in terms of the $\chi^2$-values.

The $1/N_T$ effects are much less constrained by
theoretical arguments than the $1/N_S$ effects. 
This could have resulted in a substantial
increase of the number of fit parameters necessary 
to obtain acceptable $\chi^2/N_{\mathrm{\mathrm{DF}}}$-values.
Fortunately, the $N_T$-dependence of the data is much smaller
than the $N_S$-dependence. 
We find it remarkable that, with just one additional parameter per order, we 
can accommodate the complete $N_T$ dependence down to $N_T=5$. 
Note that fitting without such an $N_T$-term to all volumes
$(\nu_T=5,\nu_S=4)$ we obtain an unacceptable $\chi^2/N_{\mathrm{DF}}=3.923$, whereas  
both choices (ii.a and ii.b) yield good reduced $\chi^2$-values,
see Table~\ref{tab:NTError}. 
Ansatz ii.b) gives results in perfect agreement with our
$\nu_T=11$ strategy, while ansatz ii.a) agrees within 1.5 standard deviations. 
In both cases we fixed the $n=0$ terms from DLPT.
We notice that the coefficients $v_n$ and $\tilde{v}_n$ are small 
in size and tend to vanish for large $n$, relative to the
divergent $c_n$ and $f_n$. 

In spite of this success, we opt for the more conservative
strategy of discarding data with $N_T<N_S$ or $N_T<11$, since
our $1/N_T$-fit ans\"atze are phenomenological and not
fully understood theoretically. For the errors associated
to the $N_T$-cut, we take the differences between 
the first columns of Tables~\ref{tab:cnCoeffs} and \ref{tab:NTError},
as this choice is completely unbiased regarding the functional form of
the $N_T$-dependence. We stress that the by far most dominant
systematics are the unknown $N_S^{-1}\ln^i(N_S)$ terms. Therefore,
alternative estimates of the $1/N_T$ effect would only marginally
affect the final errors.

We have completed the exploration of potential sources
of systematic uncertainties.
The other perturbative series (smeared, octet and octet smeared)
were analyzed analogously,
with similar conclusions and precision. In particular similar
$\chi^2$-values were obtained. The only exception was the octet case,
for which we obtained a somewhat reduced precision and the $\chi^2$-values
were smaller by factors of approximately two. This could be traced
to some geometries where the individual errors turned out
much bigger. This effect then propagated into the final data set. 

We list the final numbers for all 
the infinite volume coefficients $c_n^{(R,\rho)}$
in Table~\ref{tab:cnFinal}. The central values are taken
from the first column of Table~\ref{tab:cnCoeffs}. The quoted errors result 
from summing statistical and theoretical uncertainties in quadrature.
Schematically, we have at each order $n$
\be
\sigma_{\mathrm{final}}=\sqrt{\sigma^2_{\mathrm{stat.}}+\sigma^2_{\beta}
+\sigma^2_{T}}
\,,
\ee
where $\sigma_{\beta}$ is the difference between the first and second
columns of Table~\ref{tab:cnCoeffs}, and $\sigma_{T}$ is the difference between
the first columns of
Tables~\ref{tab:cnCoeffs} and \ref{tab:NTError}. We find
$\sigma_{\beta} \gg \sigma_{T}, \sigma_{\mathrm{stat.}}$, so 
that  the dominant error comes from logarithmic
$N_S^{-1}\ln^i(N_S)$-corrections, due to our lack of knowledge of
$\beta_3^{\mathrm{latt}}$ etc..
In comparison to these unknown $1/N_S$-terms and the
$1/N_T^d$ corrections addressed above, $1/N_S^2$ effects are negligible.

In Table~\ref{tab:cnFinal} we have chosen to multiply the
octet coefficients by factors $C_F/C_A$.
In this normalization
these will agree with the triplet coefficients for $n=0$ and $n=1$ but
at higher orders in general they will differ by $1/N_c^2$-terms.
Within our uncertainties, however, we are unable to
resolve these differences.

Our NSPT value $c_1^{(3,0)}=11.136(11)$ is in good agreement with
the DLPT expectation Eq.~\eqref{c130}. $c_2^{(3,0)}$ was calculated
previously by two groups. One group determined the static energy,
singling out the residual mass of the potential using large Wilson loops.
Employing NSPT they obtained $c_2^{(3,0)}=86.2(0.6)(1.0)$~\cite{DiRenzo:2000nd}. 
The second group fitted a polynomial in $\al$ to results of
non-perturbative simulations of the Polyakov loop at
various large values of the inverse lattice coupling $\beta$. They obtained  $c_2^{(3,0)}=86.6(5)$~\cite{Trottier:2001vj}. Our result
$c_2^{(3,0)}=86.10(13)$ confirms these studies, while
results for $n>2$ were not known previously,
e.g., $c_3^{(3,0)}=794.5(1.6)$.

The same analysis also yields the $1/N_S$ correction
coefficients $f_n^{(R,\rho)}$, where we
determine the systematic error in the same way as for the $c_n^{(R,\rho)}$.
We display the results in Table~\ref{tab:fnfinal}. 

For large orders the perturbative expansion should be dominated by
infrared physics, whereas different
smearings correspond to different regularizations of the high energy
behavior of the Polyakov loop. Therefore, we expect the smeared and
unsmeared coefficients to converge to the same values for large $n$.
This is indeed the case for the coefficients $c_n$ and $f_n$ of both the triplet and octet representations. 
Actually, the differences between smeared
and unsmeared coefficients vanish quite rapidly,
around $n=6$ for the $c_n$ and already at $n=1$ for the $f_n$.
Indeed, all smeared and unsmeared values of $f_n$ are equal within errors
for both representations. This is to be expected, as the
coefficients $f_n$ are related to finite size effects and
know nothing about the specific regularization prescription
for the ultraviolet behavior of the Polyakov loop. It is
tempting to consider global fits, constraining the
smeared and unsmeared $f_n$ values to be equal, to increase the
accuracy of the results. However, to avoid any bias
we will not explore this possibility in this article.

We now move on
to determinate the infinite volume $c_n/c_{n-1}$-ratios. These are obtained
from the same fits,
since we have also computed the correlation matrix.
Actually, we find strong correlations both of the statistical and
systematic errors between 
consecutive expansion coefficients. Due to these correlations,
the infinite volume $c_n/c_{n-1}$-ratios
can be determined more precisely
than the coefficients themselves. The results are displayed in
Table~\ref{tab:cnratioFinal}. Up to $n=11$ the errors
increase. For higher orders this tendency is reversed, since
the relative impact of the
$\beta_2$-value (and hence also
of the unknown $\beta$-function coefficients) diminishes and
so do the effects of finite $N_T$-corrections.

As a cross-check we have also determined the coefficients $c_n$ by a direct fit 
to the ratio data
\be
\frac{c_{n}}{c_{n-1}}(N_S,N_T)\Bigg|_{\mathrm{latt.}}=\frac{c_n-{f_n(N_S)}/{N_S}\quad\left[+
{v_n(N_S)}/{N_T^d}\right]}{c_{n-1}-{f_{n-1}(N_S)}/{N_S}\quad\left[+{v_{n-1}(N_S)}/{N_T^d}\right]}\,\,,
\ee
using the $f_{0,\mathrm{DLPT}}^{(R,0)}$-value for the non-smeared case
and $f_{0,\mathrm{DLPT}}^{(R,1/6)}$, obtained in the previous fit, for the
smeared case. For the central values and error estimates we proceed in the same way as 
we did before. 
 Doing this, overall consistent results and errors
for the individual coefficients are found
(with slightly bigger $\chi^2$-values). The only exception
is the unsmeared octet case, where the problems of stability
that we already encountered for the $c_n$ data become magnified in the ratios,
further reducing the precision. Subsequent
$c_n$-values are statistically correlated and 
direct fits to the ratio data take these correlations into account.
We obtain similar 
errors and central values as in the previous analysis.
This indicates that the statistical correlations of the lattice data do not significantly
affect the errors of the infinite volume coefficients, which are dominated by the systematics. 
As another related cross-check, we have computed
the infinite volume $c_n/c_{n-1}$ ratios using the ratio
data with fit parameters $c_n/c_{n-1}$, $f_n$ (and
$v_n$ or $\tilde{v}_n$, see the discussion after
Eq.~\eqref{cnNSNT}), proceeding analogously as 
above. From this we obtain very similar results to 
those quoted in Table~\ref{tab:cnratioFinal}.

Finally, we remark that at the very high orders dominated
by the renormalon behavior, TBC cannot and do not reduce finite
volume effects, relative to PBC. However, the $v_n$- and,
at low orders, the $f_n$-values are significantly reduced, considerably
increasing the robustness of the $c_n$- and $f_n$-determinations
at intermediate and large orders. The effect is twofold.
First, the impact of different parametrizations and of the
low-$N_T$ cut-off value on the $1/N_T$-extrapolation
is reduced. Second, 
the low-order $f_n$ times $\beta_3$ and higher unknown
$\beta$-function coefficients that contribute to
the $N_S^{-1}\ln^i(N_S)$-terms are smaller. Therefore,  the uncertainty due to the
lack of knowledge of $\beta_i$, $i\geq 3$, becomes reduced at
intermediate orders (for large $n$ these effects will be small
anyhow due to the renormalon dominance,
see the discussion around Eq.~\eqref{70}).

\subsection{Renormalon dominance and the determination of $N_{m}$ and
$N_{m_{\tilde g}}$}
\label{Sec:Nm}
\begin{figure}[t]
\centerline{\includegraphics[width=.9\textwidth,clip=]{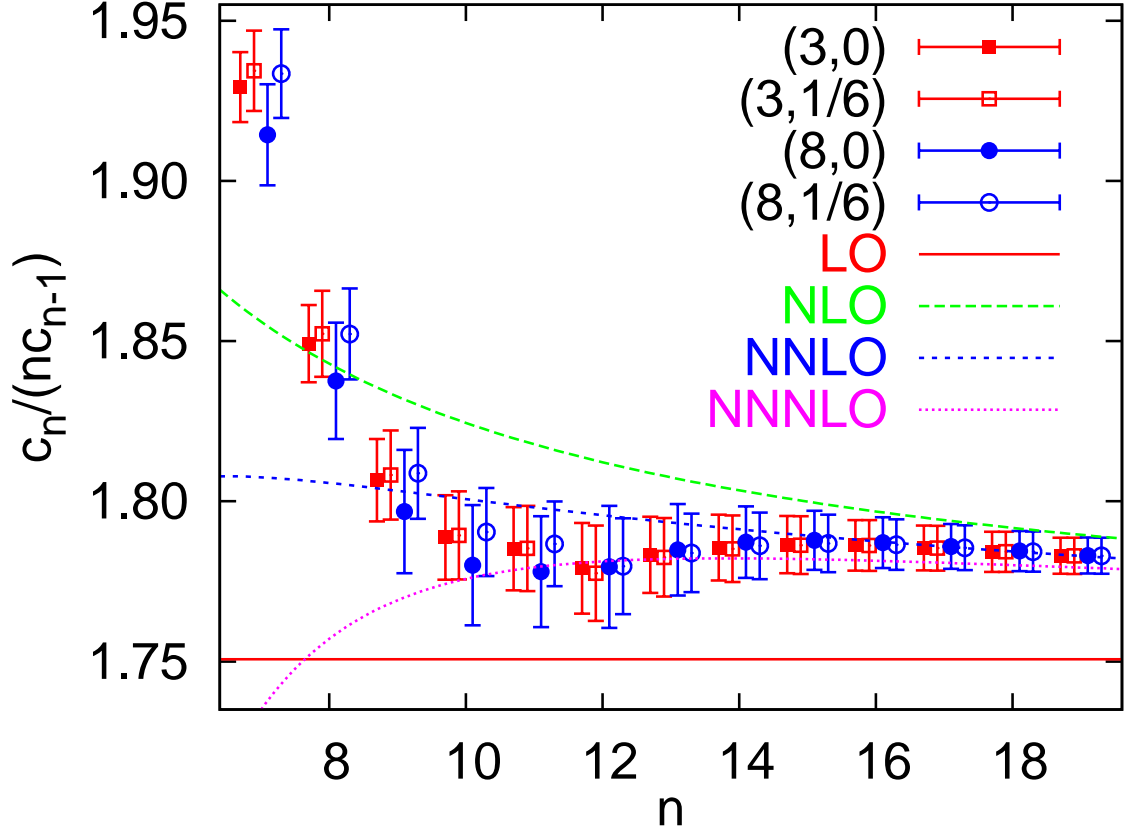}}
\caption{{\it The ratios
$c_n/(nc_{n-1})$ for the smeared and unsmeared, triplet and octet
fundamental static
self-energies, compared to the prediction
Eq.~\protect\eqref{cnratioth} for the LO,
next-to-leading order
(NLO), NNLO and NNNLO of the
$1/n$ expansion.
For clarity, the data sets are slightly shifted horizontally by different
off-sets.}
\label{n20}}
\end{figure}
In the following we investigate whether the large-$n$ behavior of the four
different sets of $c_n$ and $f_n$
complies with the renormalon expectation and determine
the triplet and octet normalizations $N_m$ and $N_{m_{\tilde{g}}}$.

In Fig.~\ref{n20} we compare the $c_n/(nc_{n-1})$-ratios summarized
in Table~\ref{tab:cnratioFinal} to Eq.~\eqref{cnratioth} at different orders 
in the $1/n$ expansion. The LO and NLO expectations are scheme independent, whereas the NNLO expression 
depends on the scheme through $\beta_2^{\mathrm{latt}}$. 
For $n\gtrsim 8$ the ratios clearly converge to Eq.~\eqref{cnratioth},
and they are within the right ball park of the NNLO 
prediction, as 
Fig.~\ref{n20} illustrates. This is so irrespectively of the representation and smearing, confirming the existence
of the renormalon at $d=1$. For completeness, we also plot
the NNNLO $\mathcal{O}(1/n^3)$ expectation, using the
$\beta_3^{\mathrm{latt}}$ estimate of Eq.~\eqref{eq:beta3}.

The renormalon picture predicts that $c_n \simeq f_n$ for large $n$. 
This equality is achieved with a high degree of accuracy
from $n=9$ onwards in all four cases (compare Tables~\ref{tab:cnFinal} and \ref{tab:fnfinal}). 
For the values of $N_S$ we explore, the renormalon picture also predicts a strong cancellation between $c_n$ and $f_n(N_S)/N_S$ for large $n$. We obtain this behavior, which we show in Fig. \ref{fig:NLrunning}, with an excellent fit to the data (see, for instance, Fig. \ref{fig:NLrunning9}, which is already at an order where renormalon dominance has set in). 

\begin{figure}
\centerline{\includegraphics[width=0.9\textwidth,clip]{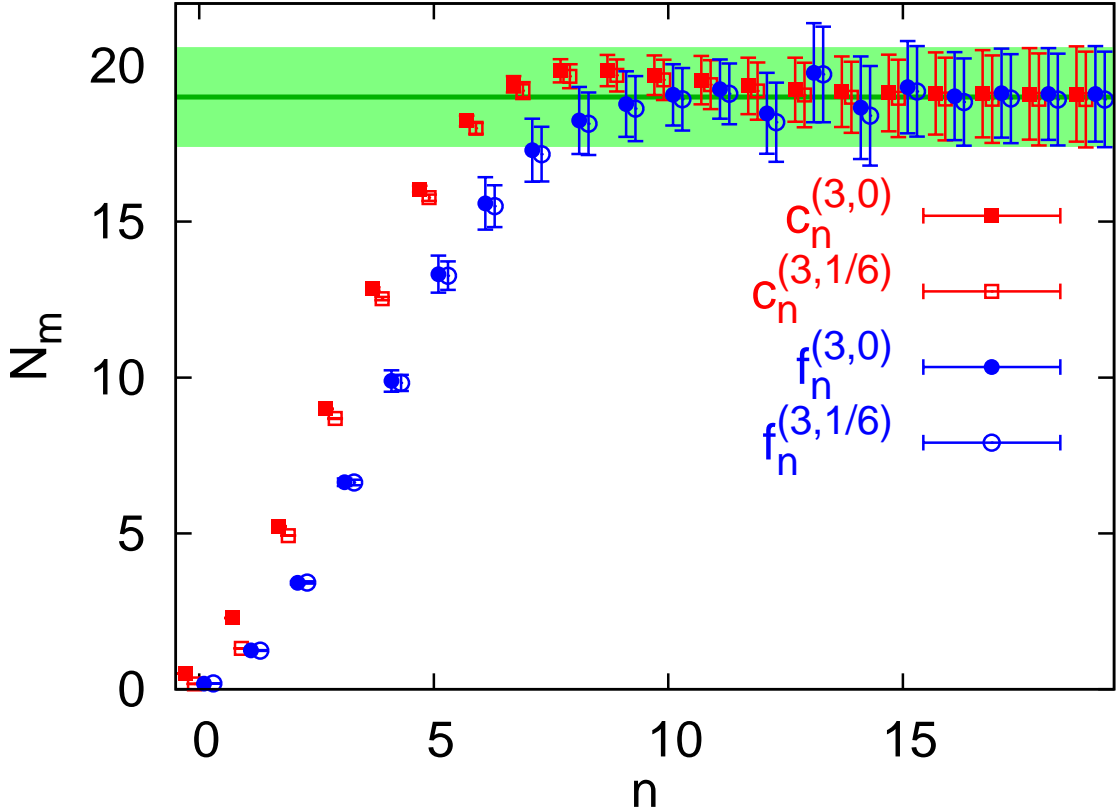}}
\caption{{\it $N_m$, determined via Eq.~\protect\eqref{generalm2},
truncated at NNLO, from the
coefficients $c_n^{(3,0)},c_n^{(3,1/6)},f_n^{(3,0)}$ and $f_n^{(3,1/6)}$. The horizontal band is our final result quoted in 
Eq.~\protect\eqref{Nmfinal}.}
\label{fig:Nm(n)}}
\end{figure}

\begin{figure}
\centerline{\includegraphics[width=0.9\textwidth,clip]{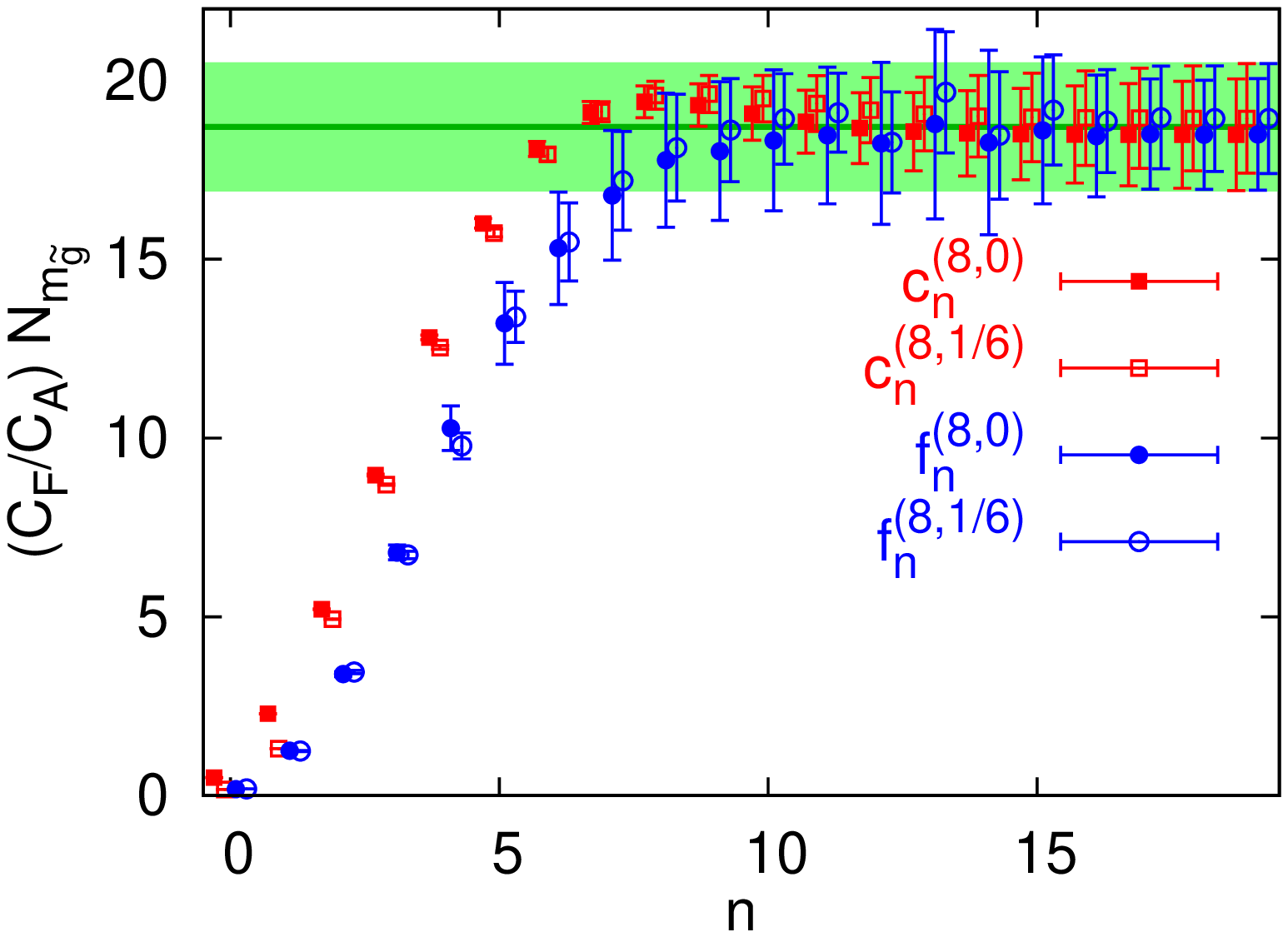}}
\caption{{\it $N_{m_{\tilde{g}}}=-N_{\Lambda}$, determined via
Eq.~\protect\eqref{cnadj},
truncated at NNLO, from the
coefficients $c_n^{(8,0)},c_n^{(8,1/6)},f_n^{(8,0)}$ and $f_n^{(8,1/6)}$. The horizontal band is our final result quoted in 
Eq.~\eqref{Nmfinal}.
To enable comparison with Fig.~\protect\ref{fig:Nm(n)},
we multiply $N_{m_{\tilde{g}}}$ by $C_F/C_A$.}
\label{fig:NLambda(n)}}
\end{figure}

For each representation $R$ we have four different
sequences: $c_n^{(R,0)}$, $c_n^{(R,1/6)}$, $f_n^{(R,0)}$ and 
$f_n^{(R,1/6)}$ that we may use to determine the normalizations
$N_m$ ($R=3$) and $N_{m_{\tilde g}}$ ($R=8$).
To obtain the normalizations we divide the large $n$ expectations
Eqs.~\eqref{generalm2} and \eqref{cnadj} for the triplet
and octet representations by 
the coefficients obtained in
Tables~\ref{tab:cnFinal} and \ref{tab:fnfinal}, respectively. We truncate the equations
at $\mathcal{O}(1/(n+b))$ precision (NNLO), since resolving the
$\mathcal{O}(1/(n+b)^2)$ correction term requires the knowledge of
$\beta_3^{\mathrm{latt}}$.
For large $n$ these ratios should tend to constants, allowing us
to extract $N_{m}$ and $N_{m_{\tilde g}}$. This is depicted in
Figs.~\ref{fig:Nm(n)} and \ref{fig:NLambda(n)} for triplet
and octet sources, respectively. We use the $n=19$ coefficients
$c_{19}^{(R,0)}$ and $c_{19}^{(R,1/6)}$, and their associated errors, 
to obtain the normalizations (recalling that $N_{m_{\tilde g}}=-N_{\Lambda}$)
\begin{align}
N^{\mathrm{latt}}_{m}(\rho=0) &= 19.1(15)\,,\quad C_F/C_A\, N^{\mathrm{latt}}_{m_{\tilde g}}(\rho=0) = 18.5(16)\ ,\\
N^{\mathrm{latt}}_{m} (\rho=1/6)&=18.9(15)\,, \quad C_F/C_A\, N^{\mathrm{latt}}_{m_{\tilde g}} (\rho=1/6)= 18.9(15)\,.
\end{align}
The errors are much bigger than the differences between the
four possible determinations: with or without smearing, using
$c_{19}$ or using $f_{19}$. This is not too surprising since these
parameters are obtained from one and the same global
fit to the same data and hence highly correlated. Moreover,
the errors are dominated by the systematics of
varying the subleading terms of the finite volume fit function.
We obtain our final result by averaging the above central
values, with errors that accommodate both the original
error bars:
\be
\label{Nmfinal}
N^{\mathrm{latt}}_{m} = 19.0\pm 1.6\, , \quad C_F/C_A\, N^{\mathrm{latt}}_{m_{\tilde g}} = 18.7\pm 1.8
\,.
\ee
These numbers are included as error bands into Figs.~\ref{fig:Nm(n)} and
\ref{fig:NLambda(n)}, respectively. The bands contain all values of
the $n\geq 8$ coefficients, lending credibility to our normalization
estimates. Note that, on general grounds, we would expect the
ratio $N_{\tilde{m}}/N_m$ to differ from the Casimir scaling
factor $C_A/C_F$ by an
$\mathcal{O}(1/N_c^2)$-term, which naively amounts to 10\%,
roughly the level of our accuracy. We discuss this issue 
further in the next subsection.

\begin{figure}[ht]
\centerline{\includegraphics[width=0.9\textwidth,clip]{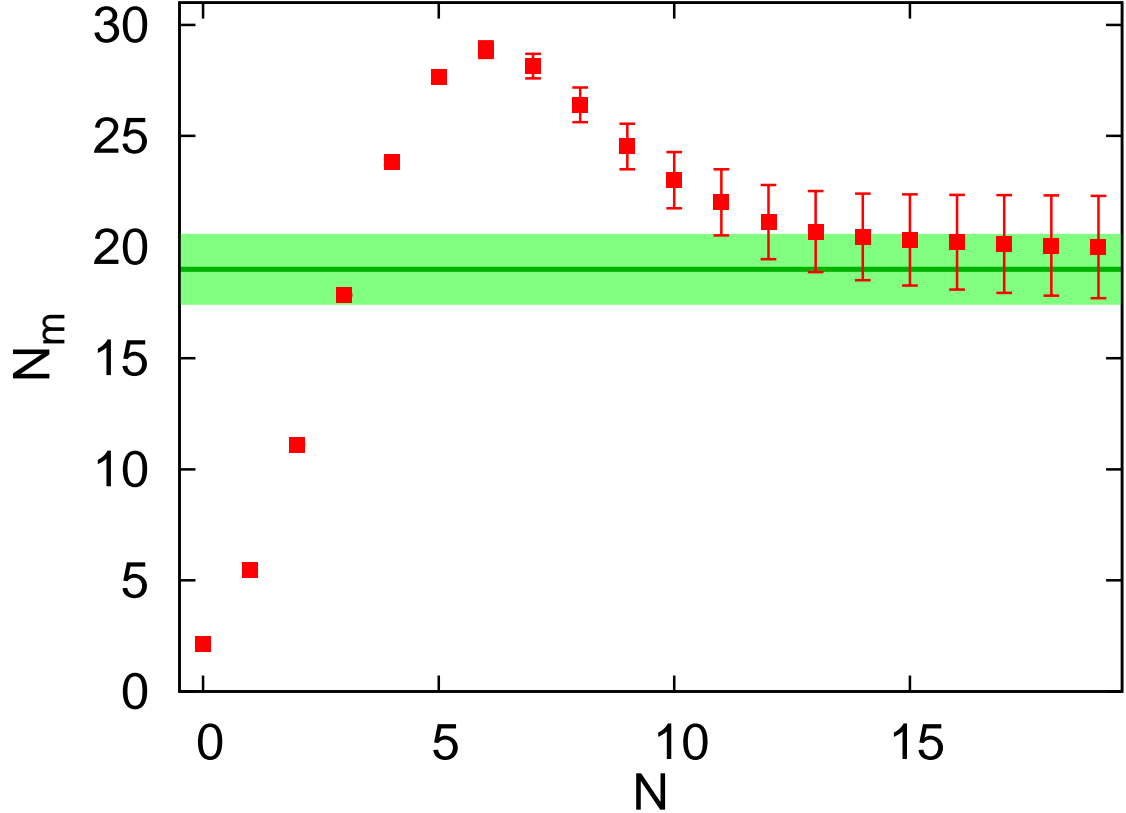}}
\caption{\it $N_{m}$, determined from the
coefficients $c_n^{(3,0)}$, $n\leq N$, using Eq.~\protect\eqref{NmDm}. The
error band corresponds to the result Eq.~\protect\eqref{Nmfinal}.
\label{fig:NmDm}}
\end{figure}

As a cross-check we also estimate the normalization
from the Borel transform of the static energy perturbative series 
\be
B^{(N)}[\delta m](t(u))= \sum_{n=0}^N
\frac{c_n}{n!}\left(\frac{4\pi}{\beta_0}u\right)^n\,,
\ee
using the function 
\begin{align}
D^{(N)}_m(u)&=\sum_{n=0}^{N}D_m^{(n)} u^n=(1-2u)^{1+b}B^{(N)}[\delta m](t(u))
\\
\nn
&
=N_m\frac{1}{a}\left(1+c_1(1-2u)+c_2(1-2u)^2+\cdots
\right)+(1-2u)^{1+b}({\rm analytic\; term})\,,
\end{align}
as it was first done in Ref.~\cite{Pineda:2001zq} for the pole mass, using ideas developed in Refs.~\cite{Lee:1996yk,Lee:1999ws}. 
$D^{(N)}_m(u)$ is singular but bounded at the first IR renormalon.
Therefore, we can estimate $N_m$ from the first
coefficients of the series in $u$, using 
\be
\label{NmDm}
N^{(N)}_m\frac{1}{a}=D^{(N)}_m(u=1/2)\,.
\ee
We plot the predictions for different orders $N$ in
Fig.~\ref{fig:NmDm}. The error is propagated from
the error of the coefficient. Within one standard deviation
the result is consistent with Eq.~\eqref{Nmfinal} (error band), though
less precise.

\begin{figure}[t]
\centerline{\includegraphics[width=.9\textwidth,clip=]{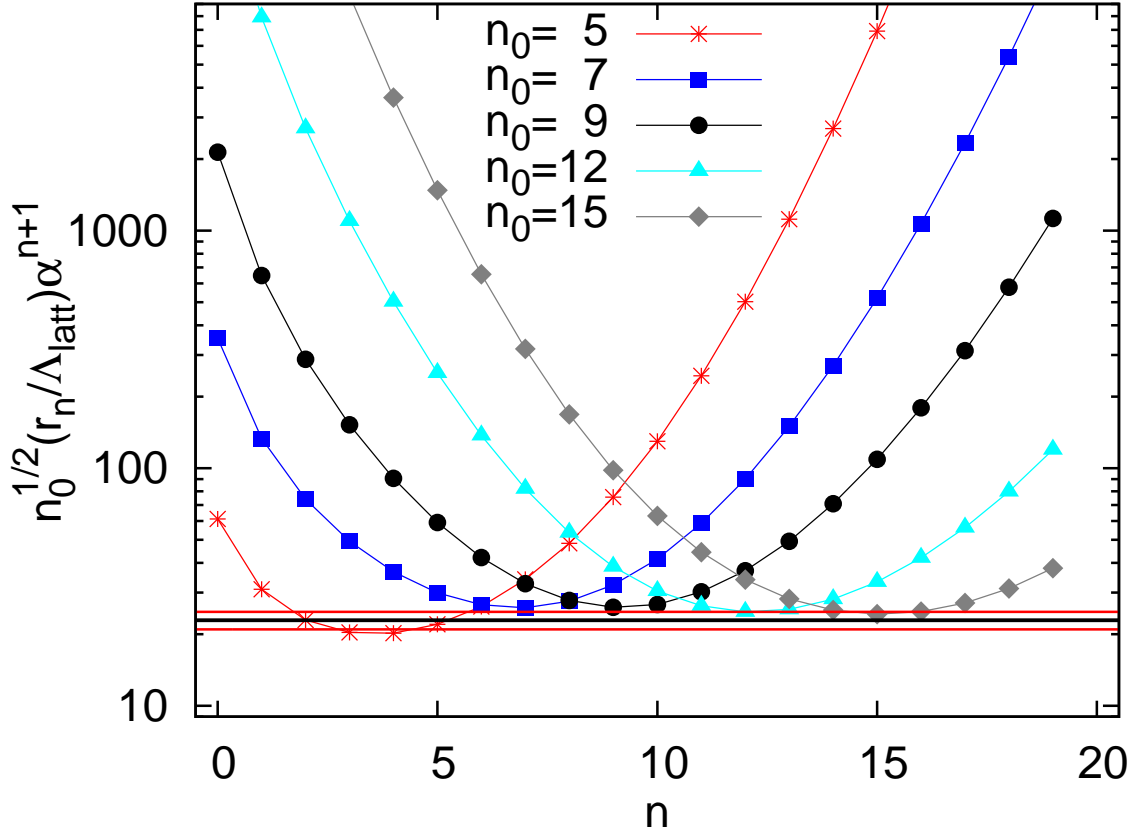}}
\caption{{\it Eq.~\protect\eqref{rnminterm} times $\sqrt{n_0}$,
for five different values of the lattice scheme coupling constant
$\alpha$, ranging from $\alpha(\nu)\approx 0.096$ ($n_0=5$)
to $\alpha(\nu)\approx 0.036$ ($n_0=15$).
The error band corresponds to the
estimate of Eq.~\protect\eqref{minterm2}, 
where we have used the value $N_m=19.0\pm 1.6$ [Eq.~\protect\eqref{Nmfinal}].}
\label{diverge3}}
\end{figure}

Finally, we show in Fig.~\ref{diverge3} the divergent behavior of the perturbative expansion of the pole mass, 
Eq.~\eqref{series}. We use the fact that $r_n=\nu c_n^{(3,0)}$ for large $n$ and 
the coefficients listed in Table~\ref{tab:cnFinal}. We compute
\be
\label{rnminterm}
\frac{r_n}{\Lambda_{\mathrm{latt}}}\alpha^{n+1}(\nu)=c_n\alpha^{n+1}(\nu)
\exp\!\left(
\frac{2\pi}{\beta_0\alpha(\nu)}\right)\left(\frac{\beta_0\alpha(\nu)}{4\pi}\right)^{\!\!b}+\cdots\,,
\ee
where we truncate Eq.~\eqref{lambdapa} at two-loop order.
In Fig.~\ref{diverge3} we plot Eq.~\eqref{rnminterm}
times $\sqrt{n_0}$ (see Eq.~\eqref{minterm2}) 
as a function of
$n$ for
 $\alpha\approx 0.096, 0.072, 0.057, 0.044$ and 0.036.
These values are chosen so that the minimal
term in the two-loop approximation of Eq.~\eqref{minimizing}
corresponds to $n_0=5, 7, 9, 12$ and 15, 
respectively. In terms of the inverse
lattice coupling parameter $\beta=3/(2\pi\alpha)$
this covers the range $4.97\lesssim\beta\lesssim 13.32$.
Orders $n_0=6,7$ ($\beta\approx 5.8, 6.6$)
are typical for present-day non-perturbative lattice
simulations, with inverse lattice spacings $1.5\,\mathrm{GeV}\lesssim
a^{-1}\lesssim 5.2\,\mathrm{GeV}$~\cite{Necco:2001xg},
while the $n_0=5$ value is in the strong coupling regime.
As expected, the contributions to the sum decrease monotonously
down to an order $\sim\alpha^{n_0+1}$, before
starting to diverge exponentially. The horizontal error band corresponds to the uncertainty,
estimated in Eq.~\eqref{minterm2},
of the sum truncated at order $n_0$
\begin{equation}
\sqrt{n_0}
\frac{|r_{n_0}|}{\Lambda_{\mathrm{latt}}}
\alpha^{n_0+1}(\nu)=
\frac{2^{3/2-b}\pi^{3/2}}{\beta_0\Gamma{(1+b)}}|N_m|\approx
1.206\,|N_m|
\,,
\end{equation}
where we used the value $N_m=19.0\pm 1.6$ [Eq.~\eqref{Nmfinal}].
Using $\Lambda_{\mathrm{latt}}\approx 8.2~\mathrm{MeV}$~\cite{Capitani:1998mq},
this horizontal line corresponds to about 190~MeV.
The data are very consistent with expectations, 
the only difference being that at the largest coupling (lowest scale
$\nu$) the order of the minimal term is somewhat lower than
expected ($n_0=3,4$ instead of $n_0=5$). We obtain
very similar results from the smeared and the octet data.

At smaller $\alpha$, i.e.\ at higher $\nu$, the
minimal term $c_{n_0}\alpha^{n_0+1}(\nu)$ is numerically smaller
than at lower scales. However, this is compensated for
by the linear divergence of $r_n=\nu c_n$, resulting
in a similar overall uncertainty.
The only difference is that to achieve this accuracy, at
higher scales one has to expand to higher orders.

\subsection{Conversion to the $\MS$ scheme and determination of $\beta_3^{\mathrm{latt}}$}

The results 
of the infinite volume coefficients $c_n^{(R,\rho)}$, $N_m$ and $N_{m_{\tilde g}}$
presented above have been obtained in the (Wilson) lattice scheme.
Translating a coefficient $c_n^{(R,\rho)}$  to a different scheme would require
the knowledge of the conversion to order $\alpha^{n+1}$.
This is completely beyond
reach. For the case of $\MS$, the conversion
\be
\label{alMSLatt}
\al_{\MS}(\mu)=\al_{\mathrm{latt}}(\mu)
\left(
1+d_1\al_{\mathrm{latt}}(\mu)+d_2\al^2_{\mathrm{latt}}(\mu)+d_3\al^3_{\mathrm{latt}}(\mu)+\mathcal{O}(\al_{\mathrm{latt}}^4)
\right)
\,,
\ee
 is known to two loops with~\cite{Hasenfratz:1980kn,Weisz:1980pu,Luscher:1995np}
$d_1=5.88359144663707(1)$ and~\cite{Luscher:1995np,Christou:1998ws,Bode:2001uz}
$d_2=43.4073028(2)$. Fortunately, only $d_1$ is needed to determine  the ratio of $\Lambda$-parameters, 
and $N^{\MS}_m$ and $N^{\MS}_{m_{\tilde{g}}}$, since (exactly!)
\be
N^{\MS}_{m,m_{\tilde{g}}}=N^{\mathrm{latt}}_{m,m_{\tilde{g}}}\Lambda_{\mathrm{latt}}/\Lambda_{\MS}\,,\quad\mbox{where}\quad
\Lambda_{\MS}=e^{\frac{2\pi d_1}{\beta_0}}\Lambda_{\mathrm{latt}}\approx
28.809338139488\,\Lambda_{\mathrm{latt}}\,.
\ee 
This yields the numerical values
\be
\label{NmfinalMS}
N^{\MS}_{m}= 0.660(56)\,, \quad C_F/C_A\, N^{\MS}_{m_{\tilde g}}=-C_F/C_A\, N^{\MS}_{\Lambda}=0.649(62)\,.
\ee
Other combinations of interest are (see Eqs.~\eqref{eq:nno1}
and \eqref{eq:nno2})
\be
N^{\MS}_{V_s}= -1.32(11) \,, \quad N^{\MS}_{V_o}=0.14(18)
\,.
\ee
These results can be compared to previous determinations from
continuum computations in the $\MS$
scheme~\cite{Pineda:2002se,Lee:2002sn,Bali:2003jq,Brambilla:2010pp}.
The agreement is remarkably
good, which is highly
nontrivial given the factor $\simeq 29$ between the values of
$N_{m}$ and $N_{\Lambda}$ in both schemes, due to the big difference
between the $\Lambda_{\MS}$- and $\Lambda_{\mathrm{latt}}$-parameters,
i.e., the large value of $d_1$. Moreover, in the $\MS$ scheme
the normalization was determined from the first few terms of the perturbative series only, while
in the lattice scheme $n\geq 9$ was required. As expected, 
the onset of the renormalon dominated behavior depends on the scheme.
Nowadays, several diagrammatic continuum perturbation theory
computations in heavy quark physics have reached a level of precision
where they become sensitive to the leading renormalon. We remark that there
has always been some doubt about the reliability of determinations
of $N^{\MS}_{m}$ and $N^{\MS}_{\Lambda}$ from just very few orders of perturbation theory.
We have now provided an entirely independent
determination of these objects based on many orders of the expansion
that can systematically be improved upon. Our quenched
result presented here goes beyond the present state-of-the-art.
An analogous un-quenched determination could give similarly precise values
for $N^{\MS}_{m}$ and $N^{\MS}_{\Lambda}$, with direct consequences to heavy quark physics,
e.g., if using the $\RS$ scheme~\cite{Pineda:2001zq}. 

To further support our conclusions,
we convert the $c_n(N_S,N_T)$ lattice coefficients, and their ratios,
into the $\MS$ scheme. 
As we have already mentioned, we can only exactly perform this
conversion up to $n=2$. 
For $n>2$ the $\MS$ coefficients and ratios will depend on the approximation used. 
If the renormalon picture is correct, the large-$n$
ratios should be dominated by the renormalon behavior and all "$\MS$-like"
conversions should yield similar results.
However, coefficients and ratios at intermediate orders will depend on the approximation
used. We consider two different $\MS$-like conversion schemes:\\
(a) $\MS_a$
\be
\label{alMSLatta}
\al_{\mathrm{latt}}(\mu)=\al_{\MS}(\mu)\frac{1}{1 + d_1 \al_{\MS}(\mu) + (d_2 - d_1^2) \al^2_{\MS}(\mu)}\ ,
\ee
(b) $\MS_b$
\be
\label{alMSLattb}
\al_{\mathrm{latt}}(\mu)=\al_{\MS}(\mu)
\left(
1-d_1\al_{\MS}(\mu)+(2d_1^2-d_2)\al^2_{\MS}(\mu)
\right)\ .
\ee
We suspect the scheme $\MS_a$ to be superior, since the translation
of $1/\al$ rather than of $\al$
from one scheme to another generates a renormalization
group-like resummation.

Our statistical data analysis~\cite{Wolff:2003sm}
allows for the direct evaluation of derived/secondary observables.
The expansion of the logarithm of the Polyakov loop is the most
obvious secondary observable and produces the coefficients $c_n(N_S,N_T)$,
but we can also intertwine the logarithm with other functions, such
as the change from the lattice to a $\MS$-like scheme. We do
so using Eqs.~\eqref{alMSLatta} and \eqref{alMSLattb}.
In addition, we employ DLPT to obtain
$c^{(3,0)}_{1,\MS}=C_F/C_A\,c^{(8,0)}_{1,\MS}=-1.3147(25)$,
whereas the first coefficient $c_0$ is scheme independent. 

\begin{figure}[t]
\centerline{\includegraphics[width=.95\textwidth,clip=]{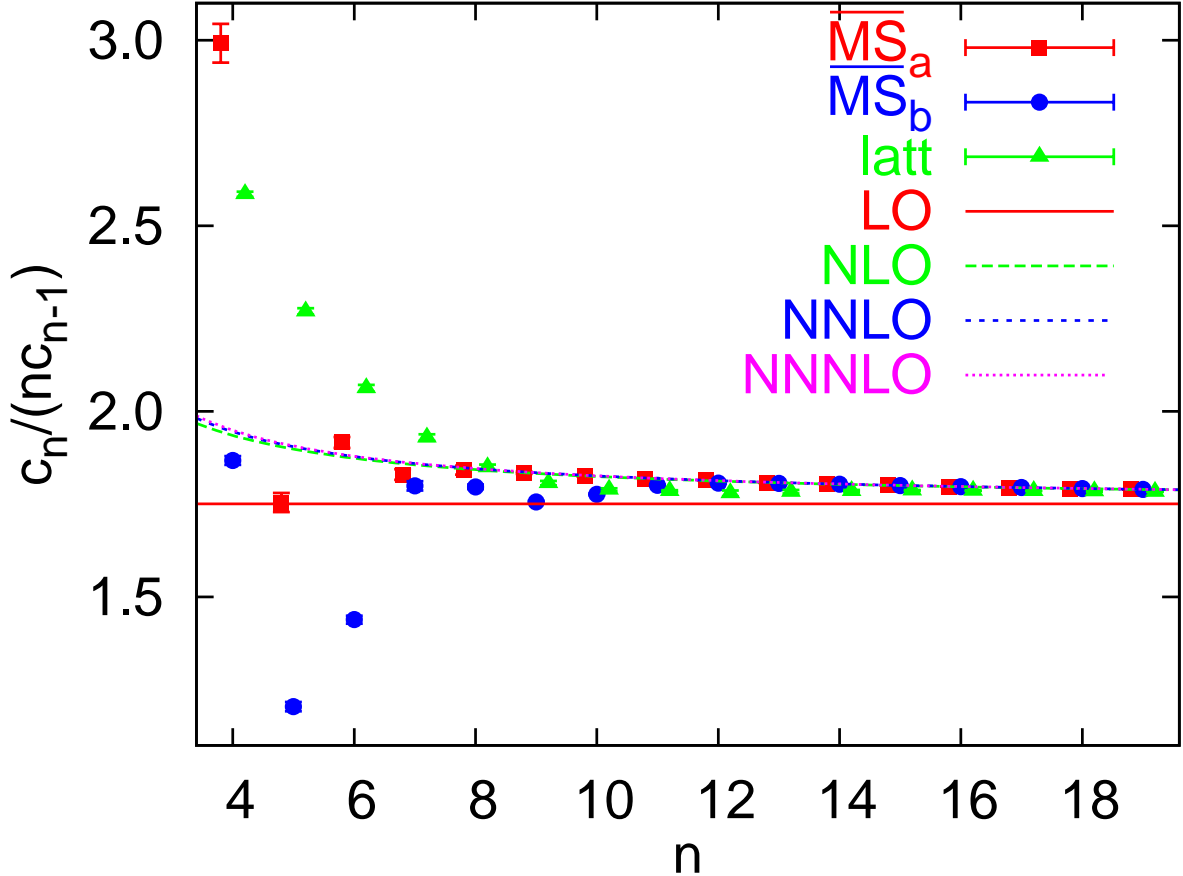}}
\caption{\it The ratio
$c^{(3,0)}_n/(nc^{(3,0)}_{n-1})$ in the lattice and $\MS$-like schemes,
compared to the prediction
Eq.~\protect\eqref{cnratioth}. NNLO and NNNLO refer to the respective
$\MS$ scheme expectations.
\label{fig:MS}}
\end{figure}

In  Fig.~\ref{fig:MS} we show our determination of $c^{(3,0)}_n/(nc^{(3,0)}_{n-1})$
using the two $\MS$-like conversions (a) and (b). We only display the statistical errors associated to the fit. We have not performed a complete error analysis, as the $\MS$-like conversions introduce unknown systematics.
As anticipated, both $\MS$-like schemes converge to the renormalon 
expectation. Actually, leaving aside systematic errors, it converges to the $\MS$ NNLO expectation rather than the lattice one. For the more stable $\MS_a$ scheme, 
renormalon dominance sets in already at orders $n\sim 5,6$, significantly earlier than in the lattice scheme. 

Following the analysis
of the Sec.~\ref{Sec:Nm}, we can determine the normalization of the renormalon
from the coefficient $c_{19}$, obtaining the estimates
\be
\label{Nmab}
N_m^{\MS_a}\simeq 0.51\ , \qquad N_m^{\MS_b}\simeq 1.53\,,
\ee
to be compared to the correct value
$N_m^{\MS}=0.66(6)$ of Eq.~\eqref{NmfinalMS}.
Of course, due to the mismatch at intermediate orders, these numbers
are not trustworthy. However, considering the
$\mathcal{O}(10^{22})$ size of $c_{19}$, the numbers
certainly are within the right ball park.
As expected, the scheme $\MS_a$ is superior, both in terms
of an earlier onset of the asymptotic behavior and the
extracted value
of the normalization $N_m^{\MS_a}\simeq N_m^{\MS}$.
A similar picture is obtained for all the other sequences except
for the unsmeared octet. In this latter case
the data become too noisy to obtain stable results.

Since we know $\beta_3^{\MS}$~\cite{vanRitbergen:1997va},
we can go one order higher in $1/n$ in the prediction for the 
ratios and coefficients of the $\MS$-like schemes. Incorporating the running to this
higher order into the fit function produces very small shifts of the 
predicted ratios and coefficients. This confirms that introducing consecutive orders
of the $\beta$-function into the fit leads
to a convergent parametrization of the $c_n$ coefficients and associated ratios.

The previous fits indicate that the asymptotic behavior
of the ratios is not very sensitive to their values at intermediate orders.
However, the normalization $N_m$ is, as the value of a
high order coefficient
$c_n$, obtained from a global fit will, through the running of the
$1/N_S$ finite size effect, also depend on $n$-intermediate orders.
This is also so if one
tries to obtain the coefficients $c_n$ through the ratios. The reason is 
that these are determined from the relation
$c_n=c_0\Pi_{j=1}^n\frac{c_j}{c_{j-1}}$, which is sensitive to the
intermediate values of $\frac{c_j}{c_{j-1}}$. 
In spite of these caveats the results are encouraging
and perfectly compatible with expectations.

We expect that
the renormalon dominance of the static energy expansion
sets in at much lower orders in the $\MS$ scheme
than in the lattice scheme. This is
supported by the consistency of our $N_m$-determination
with continuum estimates that are based on only a few orders.
Also the earlier onset of the asymptotics
in the $\MS$-like schemes is coherent with this assumption.
We can turn this argument around to
estimate $d_3$ [cf.\ Eq.~\eqref{alMSLatt}] and
$\beta_3^{\mathrm{latt}}$, assuming that 
\be
\label{c3MSassump}
c_{3,\MS} \simeq 
N^{\MS}_{m}\,\left(\frac{\beta_0}{2\pi}\right)^3
\,\frac{\Gamma(4+b)}{
\Gamma(1+b)}
\left(
1+\frac{b}{(3+b)}s_1+ \frac{b(b-1)}{(3+b)(2+b)}s_2+\cdots
\right).
\ee
Using our central value $c_{3,\mathrm{latt}}^{(3,0)}=794.5$, we obtain 
\be
\label{eq:beta3}
d_3 \simeq 365\,, \qquad \beta_3^{\mathrm{latt}} \simeq -1.7 \times 10^6\,. 
\ee
Eq.~\eqref{c3MSassump} introduces a systematic error that is
difficult to estimate. Nevertheless, we have checked that the value
of $d_3$ varies at the few per mille level when considering the
uncertainties of $N_m$, $c_{3,\mathrm{latt}}^{(3,0)}$, or when truncating 
Eq.~\eqref{c3MSassump} at a lower order in $1/n$. This translates
into variations at the level of a few per cent for $\beta_3^{\mathrm{latt}}$.
We have also checked that introducing this estimate of
$\beta_3^{\mathrm{latt}}$ in our fit function of Sec.~\ref{Sec:fitscoeff} 
yields a convergent pattern (in the number of $\beta$-coefficients included) for 
$c_n$ and $c_n/(nc_{n-1})$. In this case $c_n/(nc_{n-1})$ converges
to Eq.~\eqref{cnratioth} with NNNLO precision. The fit produces  
 a somewhat smaller value of $N_m$ that agrees
within one standard deviation with the result stated in
Eq.~\eqref{Nmfinal}.

\section{Conclusions}

We have determined the infinite volume coefficients of the
perturbative expansions of the self-energies of static sources
in the fundamental and adjoint representations to
$\mathcal{O}(\alpha^{20})$ in gluodynamics. We have employed
lattice regularization with the Wilson action and two
different discretizations of the covariant time derivative of
the Polyakov loop. The computation was performed using NSPT. 
Overall, we have obtained the infinite
volume coefficients of four different perturbative series, which we show in Table~\ref{tab:cnFinal}.
At high orders all series display the factorial growth predicted
by the conjectured renormalon picture based on the operator product
expansion. This can also nicely be seen from the normalized ratios of
subsequent coefficients $c_n/(nc_{n-1})$, which converge to
Eq.~\eqref{cnratioth} for large $n$, 
as can be read off from Table~\ref{tab:cnratioFinal}.
The coefficients that govern spatial finite size effects, $f_n$,
also grow factorially, as predicted by the renormalon dominance
picture, see Table~\ref{tab:fnfinal}. 

Furthermore, we have determined the normalization constant of the first
infrared renormalon of a heavy quark pole mass and of the gluelump mass: 
\begin{align}
N^{\mathrm{latt}}_{m} &= 19.0\pm 1.6\, , \quad C_F/C_A\, N^{\mathrm{latt}}_{\Lambda}
= -18.7\pm 1.8\, ,\\
N^{\MS}_{m}&= 0.660\pm 0.056\,, \quad C_F/C_A\, N^{\MS}_{\Lambda}=-0.649\pm 0.062\,.
\end{align}
We stress that the $N_m$-value is more than ten standard
deviations different from zero, proving, with this significance, 
the existence of the $d=1$ renormalon in gluodynamics. We also
find it remarkable that we can obtain a result in a continuum scheme
directly (and exactly) from a computation in lattice regularization,
with no error in the conversion.

The above numbers are in agreement, within errors, with determinations
from continuum-like computations, but they have been obtained using completely
independent methods. In particular, for the first
time, it was possible to follow the factorial growth of the
coefficients over many orders, from around  $\al^{9}$ up to $\al^{20}$,
vastly increasing the credibility of the prediction.
The results of this article can be used to predict higher
order terms of the heavy quark
pole mass, of the static singlet and hybrid potentials and of
the heavy gluino pole mass (gluelump) expansions.
Unfortunately, at present, for the latter we do not have
sufficient precision to discriminate Casimir scaling violation effects, 
suppressed by $1/N_c^2$ in the number of colors. 

Our precision is mainly limited by our knowledge
of the fit function, and in particular of $\beta_3^{\mathrm{latt}}$.
We have been able to estimate its value, $\beta_3^{\mathrm{latt}}\simeq-1.7\times 10^6$, 
assuming that the renormalon dominance in the $\MS$ scheme sets in
around $\mathcal{O}(\alpha^4)$. However, an independent precise determination would further 
decrease the errors of the infinite
 volume coefficients and of the normalizations $N_m$ and $N_{m_{\tilde{g}}}$. 
Performing simulations on larger lattice volumes would
also be desirable, to further improve the control of finite size effects.
However, the statistical noise increases substantially with
the length of the Polyakov loop $N_T$ and we find
simulations to become unstable for asymmetries $N_S\gg N_T$.
This behavior deserves further study. 

While the addition of a small number of quark flavors will neither
affect any of the qualitative conclusions presented here nor the renormalon
structure of the theory, a similar un-quenched analysis
would be very important. This would provide a reliable, independent
determination of $N^{\MS}_m$, including the effect of light flavors,
with major impact on renormalon analyses
in heavy quark physics and, in particular, enabling more accurate
determinations of the heavy quark masses, including that
of the top quark.

\acknowledgments{
We thank 
V.\ Braun, F.\ Di Renzo, M.\ Garc\'{\i}a P\'erez,
H.\ Perlt and A.\ Schiller for discussions.
This work was supported by the German DFG
Grant SFB/TRR-55, the Spanish 
Grants FPA2010-16963 and FPA2011-25948, the Catalan Grant SGR2009-00894
and the EU ITN STRONGnet 238353.
C.B.\ was also supported by the Studienstiftung des deutschen
Volkes and by the Daimler und Benz Stiftung.
The computations were performed on Regensburg's Athene and
iDataCool clusters
and at the Leibniz Supercomputing Centre in Munich.
}
\newpage
\appendix

\section{Tables}
\begin{table}[h]
\caption{\it Fit of $c_n^{(3,0)}$ with different approximations to
the $\beta$-function. $c^{(3,0)}_0$ and $f^{(3,0)}_0$
were fixed to the DLPT result. Leaving these parameters
free slightly decreases the $\chi^2/N_{\mathrm{DF}}$-values to 
1.111,1.152,1.177 respectively, without significant changes in any of the
fit parameters.\label{tab:cnCoeffs}}
\begin{ruledtabular}
\begin{tabular}{c|ccc}
 &$\beta_{0,1,2}$&$\beta_{0,1}$&$\beta_0$\\\hline
$\chi^2/N_{\mathrm{DF}}$&1.263 &1.290 & 1.218\\
$c_{1}/10$ & 1.1136(11)& 1.1136(11)& 1.1136(11)\\
$c_{2}/10$ & 8.610(13)& 8.610(13)& 8.597(13)\\
$c_{3}/10^2$& 7.945(14)& 7.951(14)& 7.914(14)\\
$c_{4}/10^3$& 8.215(26)& 8.232(26)& 8.156(26)\\
$c_{5}/10^4$& 9.322(40)& 9.361(40)& 9.203(40)\\
$c_{6}/10^6$& 1.1533(61)& 1.1619(61)& 1.1292(61)\\
$c_{7}/10^7$& 1.5576(96)& 1.5760(96)& 1.5067(94)\\
$c_{8}/10^8$& 2.304(16)& 2.345(16)& 2.194(15)\\
$c_{9}/10^9$& 3.747(27)& 3.837(27)& 3.499(25)\\
$c_{10}/10^{10}$& 6.702(49)& 6.913(50)& 6.121(46)\\
$c_{11}/10^{12}$& 1.3160(98)& 1.367(10)& 1.1740(89)\\
$c_{12}/10^{13}$& 2.809(24)& 2.939(24)& 2.446(21)\\
$c_{13}/10^{14}$& 6.513(56)& 6.855(58)& 5.537(51)\\
$c_{14}/10^{16}$& 1.628(14)& 1.723(15)& 1.353(13)\\
$c_{15}/10^{17}$& 4.363(38)& 4.641(40)& 3.546(33)\\
$c_{16}/10^{19}$& 1.247(11)& 1.332(11)& 0.9925(92)\\
$c_{17}/10^{20}$& 3.785(33)& 4.059(35)& 2.953(28)\\
$c_{18}/10^{22}$& 1.215(11)& 1.308(11)& 0.930(09)\\
$c_{19}/10^{23}$& 4.118(36)& 4.446(38)& 3.094(29)
\end{tabular}
\end{ruledtabular}
\end{table}
\clearpage

\begin{table}[h]
\caption{\it Determination of $c_n^{(3,0)}$ using the methods i), ii.a), and
ii.b)
explained around Eq.~\protect\eqref{cnNSNT} of
Sec.~\protect\ref{Sec:fitscoeff}. The first column is the result of a fit
with two parameters per order [Eq.~\protect\eqref{cnNS}] to the
$N_T\geq\max(N_S,9)$ geometries. The analogous $N_T\geq\max(N_S,11)$ results
are displayed in the first column of Table~\protect\ref{tab:cnCoeffs}.
The second and third columns
are from fits of Eq.~\protect\eqref{cnNSNT} to all volumes, with one
extra fit parameter per order: in the second column we set $d=2$
and obtain $v_n(N_S)$ from the renormalization group running
using $\beta_0, \beta_1, \beta_2$ and results from previous orders
$v_{n-1}^{(0)}$ etc..
In the last column we set $d=3$ and
$v_n(N_S)=\tilde{v}_n N_S$.\label{tab:NTError}}
\begin{ruledtabular}
\begin{tabular}{c|ccc}
 &$\nu_T=9$&$v_n(N_S)/N_T^2$&$\tilde{v}_nN_S/N_T^3$\\\hline
$\chi^2/N_{\mathrm{DF}}$&1.666 &0.940 & 1.033\\
$c_{1}/10$      &1.1133(10)& 1.11360(89)& 1.11442(89)\\
$c_{2}/10$      &8.607(12) & 8.612(10) & 8.619(10) \\
$c_{3}/10^2$    &7.940(12) & 7.944(10) & 7.947(10) \\
$c_{4}/10^3$    &8.201(24) & 8.233(22) & 8.231(22) \\
$c_{5}/10^4$    &9.305(37) & 9.361(34) & 9.340(35) \\
$c_{6}/10^6$    &1.1512(56)& 1.1606(52)& 1.1551(53)\\
$c_{7}/10^7$    &1.5549(88)& 1.5706(81)& 1.5589(83)\\
$c_{8}/10^8$    &2.301(14) & 2.328(13) & 2.305(13) \\
$c_{9}/10^9$    &3.742(24) & 3.791(23) & 3.745(23) \\
$c_{10}/10^{10}$&6.695(45) & 6.790(41) & 6.695(43) \\
$c_{11}/10^{12}$&1.3144(89) &1.3341(82)  & 1.3137(85)\\
$c_{12}/10^{13}$&2.812(20) &2.850(19)  & 2.805(19) \\
$c_{13}/10^{14}$&6.526(48) &6.607(44)  & 6.490(45) \\
$c_{14}/10^{16}$&1.632(12) &1.652(11)  & 1.620(11) \\
$c_{15}/10^{17}$&4.375(33) &4.426(30)  & 4.340(31) \\
$c_{16}/10^{19}$&1.2506(94) &1.2650(85)  & 1.2401(88)\\
$c_{17}/10^{20}$&3.796(28) & 3.839(26) & 3.764(27) \\
$c_{18}/10^{22}$&1.2192(92)&1.2331(83) & 1.2087(86)\\
$c_{19}/10^{23}$&4.130(31) & 4.177(28) & 4.094(29)   
\end{tabular}
\end{ruledtabular}
\end{table}
\clearpage

\begin{table}[h]
\caption{\it The infinite volume coefficients $c_n^{(R,\rho)}$, including
all systematic errors. The unsmeared $c_0$-values are fixed
using DLPT.\label{tab:cnFinal}}
\begin{ruledtabular}
\begin{tabular}{c|cccc}
 &$c_n^{(3,0)}$&$c_n^{(3,1/6)}$&$c_n^{(8,0)}C_F/C_A$&$c_n^{(8,1/6)}C_F/C_A$\\\hline
$c_0$           &2.117274357&0.72181(99)&2.117274357 & 0.72181(99)\\
$c_{1}$         &11.136(11)& 6.385(10) & 11.140(12) & 6.387(10) \\
$c_{2}/10$      &8.610(13) & 8.124(12) & 8.587(14)  & 8.129(12) \\
$c_{3}/10^2$    &7.945(16) & 7.670(13) & 7.917(20)  & 7.682(15) \\
$c_{4}/10^3$    &8.215(34) & 8.017(33) & 8.197(42)  & 8.017(36) \\
$c_{5}/10^4$    &9.322(59) & 9.160(59) & 9.295(76)  & 9.139(64) \\
$c_{6}/10^6$    &1.153(11) & 1.138(11) & 1.144(13)  & 1.134(12) \\
$c_{7}/10^7$    &1.558(21) & 1.541(22) & 1.533(25)  & 1.535(22) \\
$c_{8}/10^8$    &2.304(43) & 2.284(45) & 2.254(51)  & 2.275(45) \\
$c_{9}/10^9$    &3.747(95)  & 3.717(97)  & 3.64(11)   & 3.703(98)  \\
$c_{10}/10^{10}$&6.70(22)  & 6.65(22)  & 6.49(25)   & 6.63(22)  \\
$c_{11}/10^{12}$&1.316(52) & 1.306(53) & 1.269(59)  & 1.303(53) \\
$c_{12}/10^{13}$&2.81(13)  & 2.79(13)  & 2.71(14)   & 2.78(13)  \\
$c_{13}/10^{14}$&6.51(35)  & 6.46(35)  & 6.29(37)   & 6.45(35)  \\
$c_{14}/10^{16}$&1.628(96)  & 1.613(97)  & 1.57(10)   & 1.614(97)  \\
$c_{15}/10^{17}$&4.36(28)  & 4.32(28)  & 4.22(29)   & 4.33(28)  \\
$c_{16}/10^{19}$&1.247(86) & 1.235(86) & 1.206(89)  & 1.236(86) \\
$c_{17}/10^{20}$&3.78(28)  & 3.75(28)  & 3.66(28)   & 3.75(28)  \\
$c_{18}/10^{22}$&1.215(93)  & 1.204(94)  & 1.176(95)   & 1.205(94)  \\
$c_{19}/10^{23}$&4.12(33)  & 4.08(33)  & 3.99(34)   & 4.08(33) 
\end{tabular}
\end{ruledtabular}
\end{table}
\clearpage

\begin{table}[h]
\caption{\it The $1/N_S$ correction coefficients $f_n^{(R,\rho)}$, including
all systematic errors. The unsmeared $f_0$-values are fixed
using DLPT.\label{tab:fnfinal}}
\begin{ruledtabular}
\begin{tabular}{c|cccc}
 &$f_n^{(3,0)}$&$f_n^{(3,1/6)}$&$f_n^{(8,0)}C_F/C_A$&$f_n^{(8,1/6)}C_F/C_A$\\\hline
$f_0$           &0.7696256328& 0.7810(59)&0.7696256328& 0.7810(69)\\
$f_{1}$         &6.075(78)   & 6.046(58) & 6.124(87)  & 6.063(68) \\
$f_{2}/10$      &5.628(91)   & 5.644(62) & 5.60(11)   & 5.691(78) \\
$f_{3}/10^2$    &5.87(11)    & 5.858(76)  & 6.00(18)   & 5.946(91)  \\
$f_{4}/10^3$    &6.33(22)    & 6.29(17)  & 6.57(40)   & 6.26(23)  \\
$f_{5}/10^4$    &7.73(35)    & 7.71(26)  & 7.67(66)   & 7.78(42)  \\
$f_{6}/10^5$    &9.86(53)    & 9.80(42)  & 9.68(99)   & 9.79(69)  \\
$f_{7}/10^7$    &1.388(81)    & 1.378(71)  & 1.35(15)   & 1.38(11)  \\
$f_{8}/10^8$    &2.12(12)    & 2.11(12)  & 2.06(22)   & 2.10(17)  \\
$f_{9}/10^9$    &3.54(20)    & 3.52(20)  & 3.40(37)   & 3.51(27)  \\
$f_{10}/10^{10}$&6.49(33)    & 6.44(34)  & 6.23(67)   & 6.44(43)  \\
$f_{11}/10^{12}$&1.296(64)    & 1.286(66)  & 1.24(13)   & 1.286(74)  \\
$f_{12}/10^{13}$&2.68(19)    & 2.64(18)  & 2.65(33)   & 2.65(21)  \\
$f_{13}/10^{14}$&6.70(54)    & 6.68(52)  & 6.36(90)   & 6.66(57)  \\
$f_{14}/10^{16}$&1.58(14)    & 1.56(14)  & 1.55(22)   & 1.57(15)  \\
$f_{15}/10^{17}$&4.41(34)    & 4.37(33)  & 4.24(47)   & 4.37(35)  \\
$f_{16}/10^{19}$&1.241(92)    & 1.230(91)  & 1.20(11)   & 1.231(94)  \\
$f_{17}/10^{20}$&3.79(28)    & 3.75(28)  & 3.67(30)   & 3.76(28)  \\
$f_{18}/10^{22}$&1.215(94)    & 1.204(94)  & 1.176(97)   & 1.205(94)  \\
$f_{19}/10^{23}$&4.12(33)    & 4.08(33)  & 3.99(34)   & 4.08(33)
\end{tabular}
\end{ruledtabular}
\end{table}
\clearpage

\begin{table}[h]
\caption{\it The infinite volume ratios $c_n^{(R,\rho)}/\left(nc_{n-1}^{(R,\rho)}\right)$, including all systematic errors. Note that $\beta_0/(2\pi)\approx 1.7507$.\label{tab:cnratioFinal}}
\begin{ruledtabular}
\begin{tabular}{c|cccc}
$n$&$c_n^{(3,0)}/\left(nc_{n-1}^{(3,0)}\right)$&$c_n^{(3,1/6)}/\left(nc_{n-1}^{(3,1/6)}\right)$&$c_n^{(8,0)}/\left(nc_{n-1}^{(8,0)}\right)$&$c_n^{(8,1/6)}/\left(nc_{n-1}^{(8,1/6)}\right)$\\
\hline
1 & 5.2594(47) & 8.8462(60) & 5.2616(56) & 8.8480(61) \\
2 & 3.8662(30) & 6.3613(39) & 3.8539(36) & 6.3641(41)  \\
3 & 3.0756(41) & 3.1474(42) & 3.0735(53) & 3.1500(45) \\
4 & 2.5850(69) & 2.6129(76) & 2.5884(94) & 2.6091(79) \\
5 & 2.2695(81) & 2.2851(90) & 2.268(13) & 2.280(11) \\
6 & 2.0621(96) & 2.071(11) & 2.051(15) & 2.069(13) \\
7 & 1.929(11) & 1.934(13) & 1.914(16) & 1.933(14) \\
8 & 1.849(12) & 1.852(13)  & 1.838(18) & 1.852(14) \\
9 & 1.807(13) & 1.808(14) & 1.797(19) & 1.809(14)  \\
10 & 1.789(13) & 1.789(14) & 1.780(19) & 1.790(14) \\
11 & 1.785(13) & 1.785(13) & 1.778(17) & 1.787(13)\\
12 & 1.779(14) & 1.778(15) & 1.780(19) & 1.780(15) \\
13 & 1.783(12) & 1.782(12) & 1.785(14)  & 1.784(12)\\
14 & 1.786(10) & 1.785(10) & 1.787(11) & 1.786(10)\\
15 & 1.7865(90) & 1.7863(90) & 1.7879(92) & 1.7868(90)\\
16 & 1.7863(79) & 1.7862(79) & 1.7871(79) & 1.7865(79)\\
17 & 1.7854(70) & 1.7854(70) & 1.7859(70) & 1.7855(70)\\
18 & 1.7842(63) & 1.7842(63) & 1.7845(62) & 1.7843(63)\\
19 & 1.7830(56) & 1.7830(56) & 1.7831(56) & 1.7830(56)
\end{tabular}
\end{ruledtabular}
\end{table}

\end{document}